\documentclass[letterpaper, 11pt]{article}
\usepackage[T1]{fontenc}
\usepackage{lmodern} 
\usepackage{amsmath}
\usepackage{amsfonts}
\usepackage{fancyhdr}
\usepackage{float}
\usepackage[labelfont=bf,font=footnotesize]{caption}
\usepackage{subcaption}
\usepackage{graphicx}

\usepackage[english]{babel} 
\usepackage[colorinlistoftodos]{todonotes}
\usepackage{comment}

\makeatletter
\def\toclevel@paragraph{4}
\def\l@paragraph{\@dottedtocline{4}{10em}{5em}}
\def\l@subparagraph{\@dottedtocline{5}{14em}{6em}}
\makeatother

\usepackage[dvipsnames,table]{xcolor}
\definecolor{blue600}{RGB}{0,136,181}
\definecolor{blue900}{RGB}{0,69,115}
\definecolor{orange600}{RGB}{217,99,0}
\definecolor{green600}{RGB}{0,150,10}
\definecolor{slate100}{RGB}{195,199,217}
\definecolor{slate800}{RGB}{77,81,94}
\definecolor{black700}{RGB}{61,61,61}
\definecolor{darkgreen}{rgb}{0,0.5,0}
\definecolor{orange}{rgb}{0.7,0.5,0.0}
\definecolor{brown}{rgb}{0.6,0.3,0.0}

\usepackage[letterpaper, margin=2cm]{geometry}
\usepackage{abstract}
\usepackage{booktabs}

\usepackage[hidelinks]{hyperref}

\usepackage{CormorantGaramond} 

\newif\ifdraft
\draftfalse

\ifdraft
\usepackage[firstpageonly=true]{draftwatermark}
\SetWatermarkText{DRAFT}
\newcommand{\cg}[1]{{\it \color{purple} CG: #1 --}}
\newcommand{\ag}[1]{{\small\it \color{magenta} AG: #1 --}}
\newcommand{\oli}[1]{{\it \color{darkgreen} Oli: #1 --}}
\newcommand{\yak}[1]{{\it \color{orange} YAK: #1 --}}
\newcommand{\john}[1]{{\it \color{blue} John: #1 --}}
\newcommand{\TODO}[1]{\textcolor{red}{ TO DO: #1 --}}
\newcommand{\assignedto}[1]{\textcolor{red}{[Assigned to: #1 ]}}
\newcommand{\assignedspace}[1]{\textcolor{red}{[Assigned space: #1 pg]}}
\newcommand{\instruction}[1]{\color{brown}{\par \textbf{Instruction:} #1 \par} \color{black}}
\newcommand{\inprogress}[1]{\textcolor{magenta}{[In progress: #1 ]}}
\newcommand{\due}[1]{\textcolor{magenta}{[Due: #1 ]}}
\newcommand{\ALT}[1]{{\color{Violet} #1 - {\bf Alternate Text}}}
\newcommand{\preprop}[1]{{\color{cyan} {\bf begin quote from Preproposal, needs  rewrite}} - #1 - {\color{cyan} {\bf end quote from Preproposal, needs  rewrite}} \\ }
\else
\newcommand{\cg}[1]{}
\newcommand{\ag}[1]{}
\newcommand{\oli}[1]{}
\newcommand{\yak}[1]{}
\newcommand{\john}[1]{}
\newcommand{\TODO}[1]{}
\newcommand{\assignedto}[1]{}
\newcommand{\assignedspace}[1]{}
\newcommand{\instruction}[1]{}
\newcommand{\inprogress}[1]{}
\newcommand{\due}[1]{}
\newcommand{\ALT}[1]{}
\newcommand{\preprop}[1]{}
\fi

\title{\textbf{\color{orange600} Building AI That Works: ESnet's Pragmatic Approach to
  AI-Driven Operational Excellence}\\
  \textcolor{BrickRed}{-- A progress report on developing Operations
    Response \& Business Intelligence Toolkit (ORBIT)}}

\author{\textcolor{blue600}{Bin Dong, Sukhada Gholba,  Brooklin Gore, Shawn Kwang,} \\
    \textcolor{blue600}{David Mitchell, Samuel Oehlert, Garrett Stewart,  Brendan White,} \\
    \textcolor{blue600}{Luke Baker, Ed Balas, Britt Gathright, Chin Guok,} \\
    \textcolor{blue600}{Jon-Paul Heron, John MacAuley, Scott Richmond,} \\
    \textcolor{blue600}{Chris Robb, Chris Tracy, Kesheng Wu} \\ \\
\textcolor{black700}{Energy Science Network} \\
\includegraphics[width=3cm]{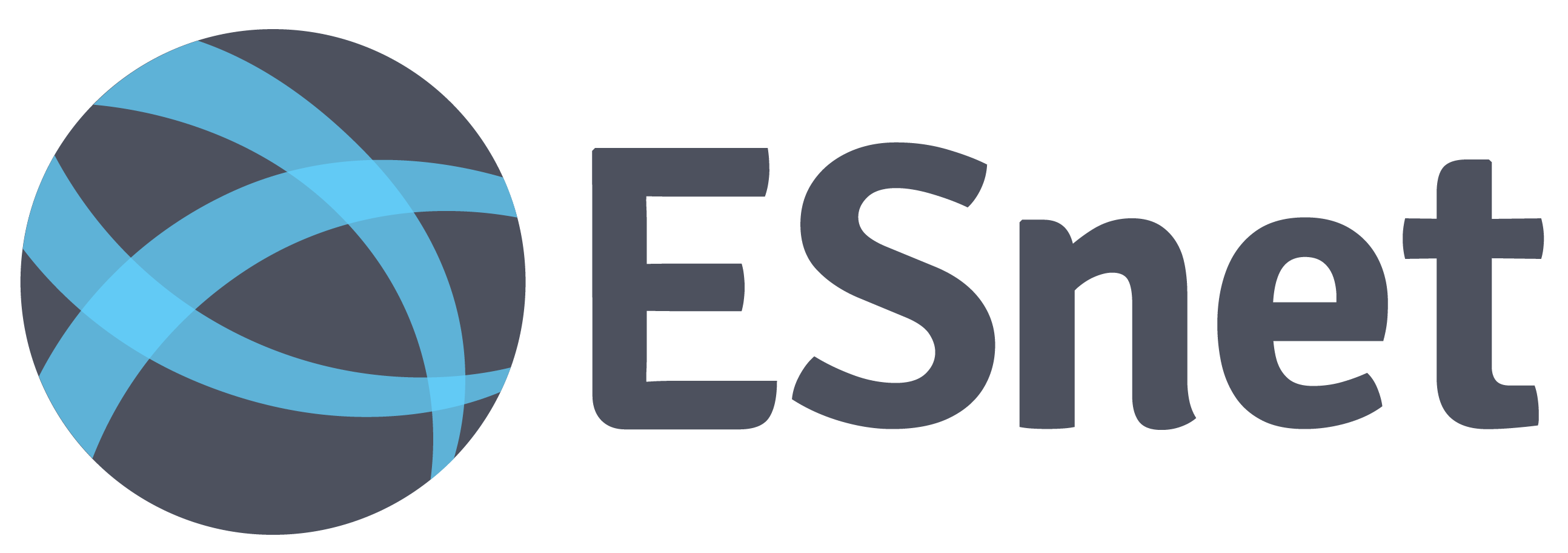}
}

\date{June 2026} 

\addto\captionsenglish{

}

\begin{document}
\pagenumbering{gobble} 
\twocolumn[
\maketitle 
\begin{onecolabstract}
\vspace{-1.5em}
The ORBIT (Operations Responses \& Business Intelligence
Toolkit) project was initiated to explore the potential of agentic AI for
upcoming ESnet 7 initiative.  This report summarizes the project's
objectives, design, key findings, and recommendations, providing a
comprehensive overview of the operational and strategic value of AI-driven
tools in a production environment.

ORBIT is grounded with a present business needs to address persistent pain
points in the NOC workflow.  These include slow information retrieval from
siloed data sources, lengthy and difficult-to-parse incident tickets, and
context loss during shift handoffs.  These challenges result in increased
cognitive load on operators and longer incident resolution times.  To
address these issues, the ORBIT project set out to develop an AI-powered
system that could automate routine tasks, synthesize information from
multiple sources, and provide operators with actionable insights directly
within their existing workflows.

ORBIT is an agentic AI system integrated into ServiceNow, the NOC's primary
incident management platform.  The system is built on a modular, layered
architecture that includes a centralized reasoning hub, a set of specific
tools (MCPs) for accessing ESnet data sources, a semantic search layer, and
a chat interface.  The ORBIT project has adopted the industry best practice
of using ``skills'' to manage the complexity and stochasticity of the AI
toolchain.  Skills are versioned, tested, and iteratively refined sets of
instructions that guide the AI in performing specific tasks, ensuring
greater reliability and predictability.

The ORBIT project has yielded several important findings.  The system
successfully delivered on all six of its initial tasks, and its
flexible architecture enabled the rapid development of two additional tasks
proposed by NOC engineers.  We observed significant organic adoption of the
project's general-purpose infrastructure components, with the chat interface
and LiteLLM model gateway attracting a large number of users and a high
volume of requests from outside the project.  This demonstrates a clear
demand for accessible, well-supported AI tools within ESnet.  Furthermore,
our experiments with ``skills'' have shown that this approach can
dramatically improve the performance and reliability of the AI, reducing the
number of steps required to complete a task and eliminating errors.

Based on these findings, we recommend that ESnet adopt the four
general-purpose components of ORBIT as officially supported services.  This
will require a dedicated productization effort to establish clear ownership,
governance, and support structures.  We also recommend continued investment
in a comprehensive skill library to further improve the reliability and
effectiveness of the system.  Additional future work may include a
systematic evaluation of mission-oriented outcomes, the development of a
unified, identity-aware tool gateway, and a continued focus on data
governance and operational sustainability.

\end{onecolabstract}
]

\clearpage

\pagenumbering{arabic} 
\tableofcontents

\centerline{\rule{0.45\columnwidth}{0.5pt}}

\section{Introduction} \label{src:intro}

\begin{figure*}
  \centerline{\includegraphics[width=0.75\textwidth]{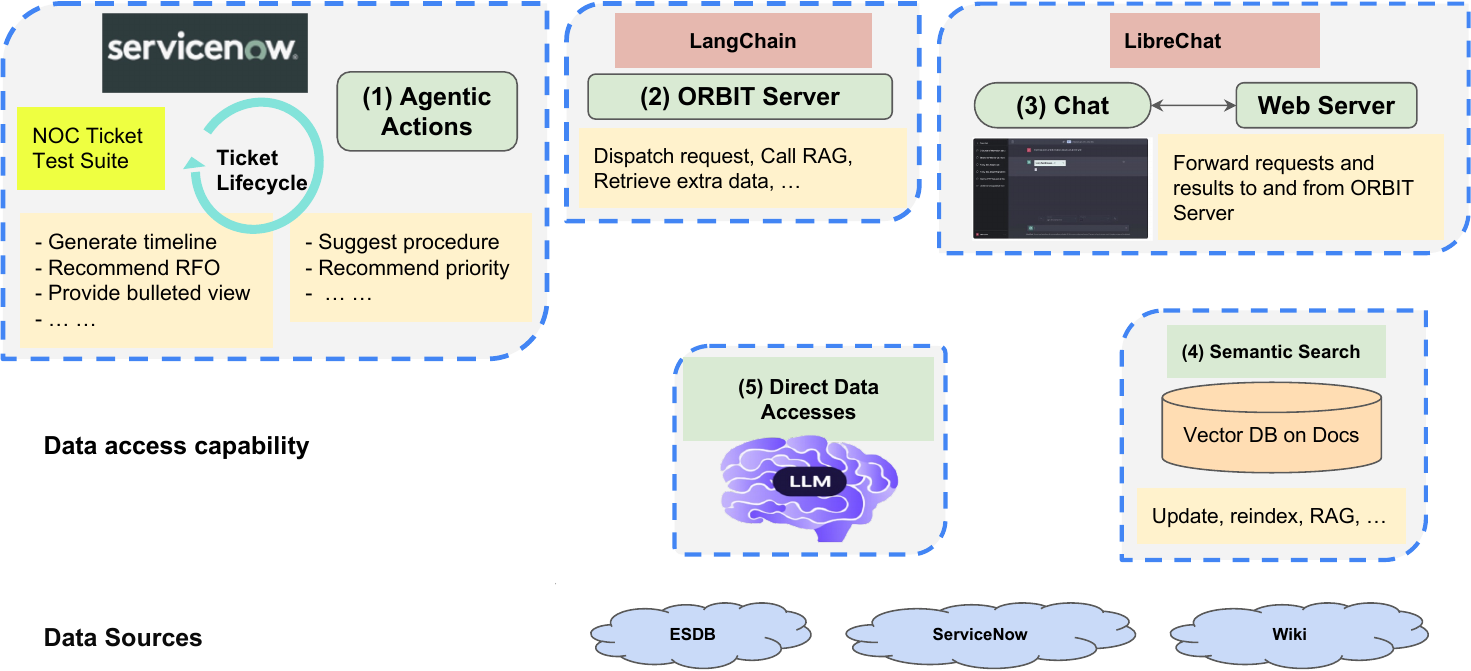}}
  \caption{A logical view of ORBIT where the key NOC functions sit on top of
    a set of tools that integrate data from multiple sources.}
  \label{fig:logical-view}
\end{figure*}

\begin{table*}
  \caption{List of NOC tasks targeted for the ORBIT project.}
  \label{tb:tasks}
  \begin{tabular}{|l|p{1.1in}|p{0.9in}|p{3.5in}|} \hline
  & \textbf{Activity} & \textbf{Data} & \textbf{Description} \\ \hline

1 & Summarize \newline incident \newline for hand-off & ServiceNow & When a ticket is handed
off to another team or escalated, parse the ticket and generate a history of
the actions taken so far in a summary that is easy for the next person to
process. \\ \hline

2 & Generate ticket \newline timeline & ServiceNow, \newline ESDB & On demand,
generate a timeline of the ticket with all actions taken and resolution
information for inclusion in an after action report. \\ \hline

3 & Suggest alarm \newline procedure &
ESDB, \newline ESnet Wiki, \newline Stardust, \newline ServiceNow &
Place all the NOC process documentation behind AI and use the alarm data in
a given ticket to automatically populate it with relevant potential next
steps from our standard processes when the ticket is created. Place this
information inline in the ticket with references to the documentation
cited. \\ \hline

4 & Recommend \newline alarm priority &
ESnet Wiki, \newline ESDB, \newline ServiceNow & Use
NOC process documentation to recommend any changes to the default alarm
priority given the alarm profile and impact. \\ \hline

5 & Propose final\newline RFO &
ESDB, \newline ESnet Wiki, \newline ServiceNow & Using NOC process
documentation and our RFO codes, process the ticket narrative and determine
a suggested RFO code and close note summary. \\ \hline

6 & NOC chat bot &
ESnet Wiki, \newline ESDB &
Place all the NOC process documentation
behind AI and create a chat bot that enables a user to ask questions of our
documentation along with references to the source documentation. \\ \hline
\end{tabular}
\end{table*}

While advances in large language models make AI assistance increasingly
accessible, deploying AI in network operations demands more than model
novelty; it requires measurable improvements to operational workflows under
realistic constraints~\cite{Acharya2024NetworkOperations,
  Bughin2019Transformation, Cooper2024AI, Raza2025industrial,
  Westenberger2022AI}. Research and engineering groups managing
high-performance research networks face recurring challenges in incident
handling: finding the right resolution context across heterogeneous
knowledge sources, communicating status to multiple audiences, and
maintaining operator control over AI outputs~\cite{Guok:2025:ESnetData,
Long2025AIops, Min2024AIops, Pan:2021:Net+AI, Westenberger2022AI}.  These
challenges are compounded by the inherent stochasticity of AI tool chains, which
can lead to unpredictable and unreliable behavior.  These
requirements motivate a shift from benchmarking model capabilities in
isolation to considering business needs, technology trends, and staff
engagement in production settings~\cite{Rajan2025pilots, Shaik2026AI,
  Vallone2025AI}.

Our work began with a systematic review of business
needs~\cite{Guok:2025:ESnetData}, followed by a careful consideration of
available resources and technical expertise~\cite{Powell:2025:AIData}.
We then selected a set of tasks from ESnet's Network Operation Center (NOC)
workflow to drive a data integration and platform development effort (see
\autoref{fig:logical-view} for an illustration). Following this decision,
the planning team worked with engineering staff to develop concrete tasks
from NOC operations. Based on considerations such as data availability,
staff availability, task impact, and task feasibility, the team selected
the six specific NOC tasks listed in \autoref{tb:tasks}.

The project had a limited duration (January through June 2026) with three
primary objectives: (1) develop the Operations Responses \& Business
Intelligence Toolkit (ORBIT) to accomplish the six specific tasks, (2)
engage ESnet staff in exploring AI tools, and (3) gather performance and
usability statistics to inform future AI efforts. This report summarizes the
ORBIT project, detailing our software system, usage observations, and
system performance measures in support of all three project goals.

ORBIT is a pragmatic AI system integrated into a Network Operation Center
(NOC) workflow to assist with incident handling. Our design is
stakeholder-driven and developed through incremental deployment, guided by
an explicit goal: augment operator capabilities without introducing
additional process overhead. ORBIT is built to fit existing operational
 tooling, most notably ServiceNow incident work notes, while preserving
 human-in-the-loop decision-making throughout.

This work makes three contributions, one aligned with each project
objective:

\begin{enumerate}
\item \textbf{A workflow-integrated agentic system for NOC incident
  handling.} ORBIT delivers two categories of production-oriented
  capability. Adaptive Incident Summarization (Tasks~1--2 in
  \autoref{tb:tasks}) generates audience-specific summaries from ServiceNow
  incident narratives to streamline handoffs and stakeholder communication.
  Intelligent Resolution Guidance (Tasks~3--5) retrieves and ranks
  resolution procedures from operational knowledge bases using incident
  context, surfacing relevant steps and supporting evidence when operators
  would otherwise rely on manual documentation search. Both capabilities
  are grounded in cross-source synthesis across ServiceNow, Confluence,
  Slack, ESDB, and Stardust via purpose-scoped MCP tools and hybrid
  semantic search integrated directly into the ServiceNow incident
  workflow. Beyond the six originally scoped tasks, the platform's
  composability enabled two additional tasks---Caused-By-Change Analysis
  and Change Analysis---proposed by NOC engineers and implemented within
  hours.

\item \textbf{Reusable AI infrastructure that attracted adoption beyond
  the original project scope.} Four components---(1) the LiteLLM model
  gateway, (2) the agentic application service, (3) the chat interface
  (\texttt{chat.es.net}), and (4) the MCP tool layer---were designed as
  general-purpose services rather than project-specific utilities. The
  agentic service exposes a uniform API through which any client can
  compose task-specific workflows from pluggable tools, versioned prompt
  templates, and selectable language models, making it reusable as a
  general-purpose agentic composition service beyond incident handling.
  During the evaluation period, the gateway served over 714,000~requests
  consuming more than 21~billion tokens, of which roughly 95\% came from
  coding-assistant workloads initiated by engineers outside the ORBIT
  project. Six of seven NOC operators adopted the chat interface
  organically, generating 72~conversations and over 1,200~tool
  invocations without prescribed workflows---substantially more
  interaction than the 12~user-initiated ServiceNow AI Actions recorded
  in the same period. This infrastructure realizes the AI Sandbox called
  for in ESnet's Data and AI report~\cite{Guok:2025:ESnetData}.

\item \textbf{Operational measurements and engineering lessons from a
  test deployment.} We report per-task execution statistics,
skill-refinement evidence (engineered skills reduced agent actions
from 10 to~4 and eliminated retries), workload-level cost breakdowns,
and structured user interviews with NOC operators. Five engineering
lessons distilled from this evidence address skill versioning and
testing, stochasticity in tool selection, emergent use cases, organic
  infrastructure adoption, and the cognitive barriers to operational AI
  adoption. Together, these measurements and lessons illustrate how
  operational AI systems can be evaluated against workflow requirements
  rather than relying on benchmark scores or subjective impressions of
  helpfulness.
\end{enumerate}

\section{Background} \label{sec:bg}
\begin{figure}
  \centerline{\includegraphics[width=0.48\textwidth]{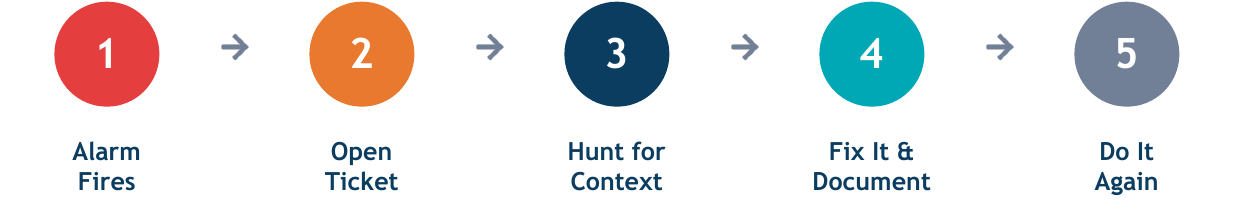}}
  \caption{A schematic drawing of a simplified NOC workflow.}
  \label{fig:noc-workflow}
\end{figure}

Operational teams increasingly adopt AI assistance to reduce the time and
cognitive load associated with incident response~\cite{Veluru:2021:Incident,
  Joy2024AIops}.  In practice, success depends less on demonstrating model
novelty and more on achieving measurable improvements in operator workflows
under real-world constraints: heterogeneous data sources, variable incident
narratives, strict governance requirements, and the need for
human-in-the-loop decision-making~\cite{Acharya2024NetworkOperations,
  Bughin2019Transformation, Cooper2024AI, Raza2025industrial,
  Westenberger2022AI}.  Network Operations Centers (NOCs) face
recurring bottlenecks in context gathering and communication.  When alarms
trigger, operators must quickly determine what happened, identify the
relevant components and procedures, and translate ticket histories into
accurate status updates for different stakeholders.  Much of this effort is
spent searching across disjoint knowledge bases rather than executing
resolution steps (see \autoref{fig:noc-workflow}).

In practice, the pain concentrates in four areas that compound one another:

\begin{itemize}
    \item \textbf{Finding information is slow.} Operators search across
      documentation, past tickets, and Slack conversations to locate the
      relevant procedure, often re-finding the same material they located
      weeks earlier.  The absence of a unified retrieval path turns every
      new alarm into a repetitive, time-consuming scavenger hunt.

    \item \textbf{Tickets become novels.} ServiceNow incidents accumulate
      pages of work notes, comments, and status updates over their
      lifetimes.  Understanding the state of an incident means
      reading everything; there is no reliable shortcut for extracting the
      salient facts from a lengthy narrative.

    \item \textbf{Handoffs lose context.} When tickets move between shifts
      or are escalated to another team, the receiving operator effectively
      starts from scratch.  Critical context---what was tried, what was
      ruled out, what the customer was told---is embedded in the ticket
      narrative but not surfaced in a form that supports rapid onboarding.

    \item \textbf{Knowledge is everywhere and nowhere.} Procedures live in
      Confluence, real-time context lives in Slack, incident history lives
      in ServiceNow, and equipment details live in ESDB.  These sources are
      almost never consulted together, forcing operators to mentally stitch
      together a coherent picture from fragments scattered across systems.
\end{itemize}

At ESnet, these challenges motivated the exploratory project
ORBIT~\footnote{Some earlier documents refer to this as ``Ops Assistant.''}.
The project originated from ESnet's internal review of high-impact AI use
cases, emphasizing workflows that are time-consuming, difficult to execute
consistently, and feasible to integrate into operational
tooling~\cite{Guok:2025:ESnetData, Powell:2025:AIData}.  Through
collaboration with NOC operators, we narrowed an initial set of candidate
use cases down to six that share a common theme: operators already possess
the necessary knowledge, but it is buried across multiple systems and
formats.  The four pain points above map directly onto the ORBIT task set:
Adaptive Incident Summarization addresses the ``tickets are novels'' and
``handoffs lose context'' problems by generating concise,
audience-appropriate summaries, while Intelligent Resolution Guidance
targets the ``finding information is slow'' and ``knowledge is everywhere''
problems by retrieving and ranking relevant procedures from operational
knowledge bases using incident context.  Accordingly, ORBIT targets these
two categories of functionality (see \autoref{tb:tasks}).

A key driver for this work is data integration.  AI tools are known to
be well-suited for integrating data from multiple
sources~\cite{althati2024enhancing, Lipkova2022Integration, Reddy2025Integration}, and a
unified data access mechanism is needed for more than half of the
work-packages identified from our workshop~\cite{Guok:2025:ESnetData, Powell:2025:AIData}.  The tasks in
\autoref{tb:tasks} exercise a small set of input sources: the NOC
process and procedure documentation captured as ESnet Wiki pages in
Confluence (``Wiki/Confluence data''), ServiceNow incident reports
(``tickets''), Stardust for ESnet telemetry~\cite{Balas:2022:Stardust}, and
ESDB (ESnet Database) for physical and logical attributes of network
components.  We limited the number of sources to ensure the
evaluation would primarily characterize data integration and synthesis,
while providing coverage of realistic operational content.

A fundamental design constraint is that AI assistance must integrate
seamlessly into the ServiceNow-centric operational workflow without creating
additional overhead or shifting accountability away from operators.
Consequently, ORBIT is explicitly not designed to autonomously decide or
execute resolution actions.  Instead, it generates evidence-grounded drafts
and suggestions inserted into ticket work notes and editable fields,
enabling operators to accept, refine, or reject AI output.  This supports
governance and also enables evaluation: operator decisions and edits provide
observable signals of usefulness and failure modes.

Another key motivation is measurability.  Rather than relying on subjective
impressions of ``helpfulness,'' the project requires instrumentation
supporting usability and performance evaluation for deployment readiness and
broader investment decisions.  ORBIT, therefore, collects operationally
relevant data such as LLM call counts, latency, and resource consumption
(with tokens used as a proxy for cost), along with user feedback gathered
from the same interface used by operators.  A guiding principle is that
software capability alone is insufficient: the team must gather enough
evidence to characterize the operational impact and inform whether ORBIT should
proceed to broader use or similar AIOps initiatives.  Stakeholder feedback also identified evaluation pitfalls,
including contamination of tickets when AI actions are repeated.  This drives
requirements for mechanisms to separate AI-generated content from original
 a ticket text so that performance measurement remains interpretable.

Finally, ORBIT serves as ESnet's initial operational test deployment of AI
components.  Beyond improving incident handling, the project aims to build
organizational capability for designing, integrating, and governing AI
systems in production.  This includes security evaluation under ESnet's AI
and cloud policies, careful handling of business-sensitive network
configuration data in ESDB, and structured cross-department collaboration to
establish a foundation for responsible AI operations via both technical
integration patterns and an evaluation methodology grounded in
operational measurements.

\section{System Overview} \label{sec:sys}

\begin{figure*}
\centerline{\includegraphics[width=0.6\textwidth]{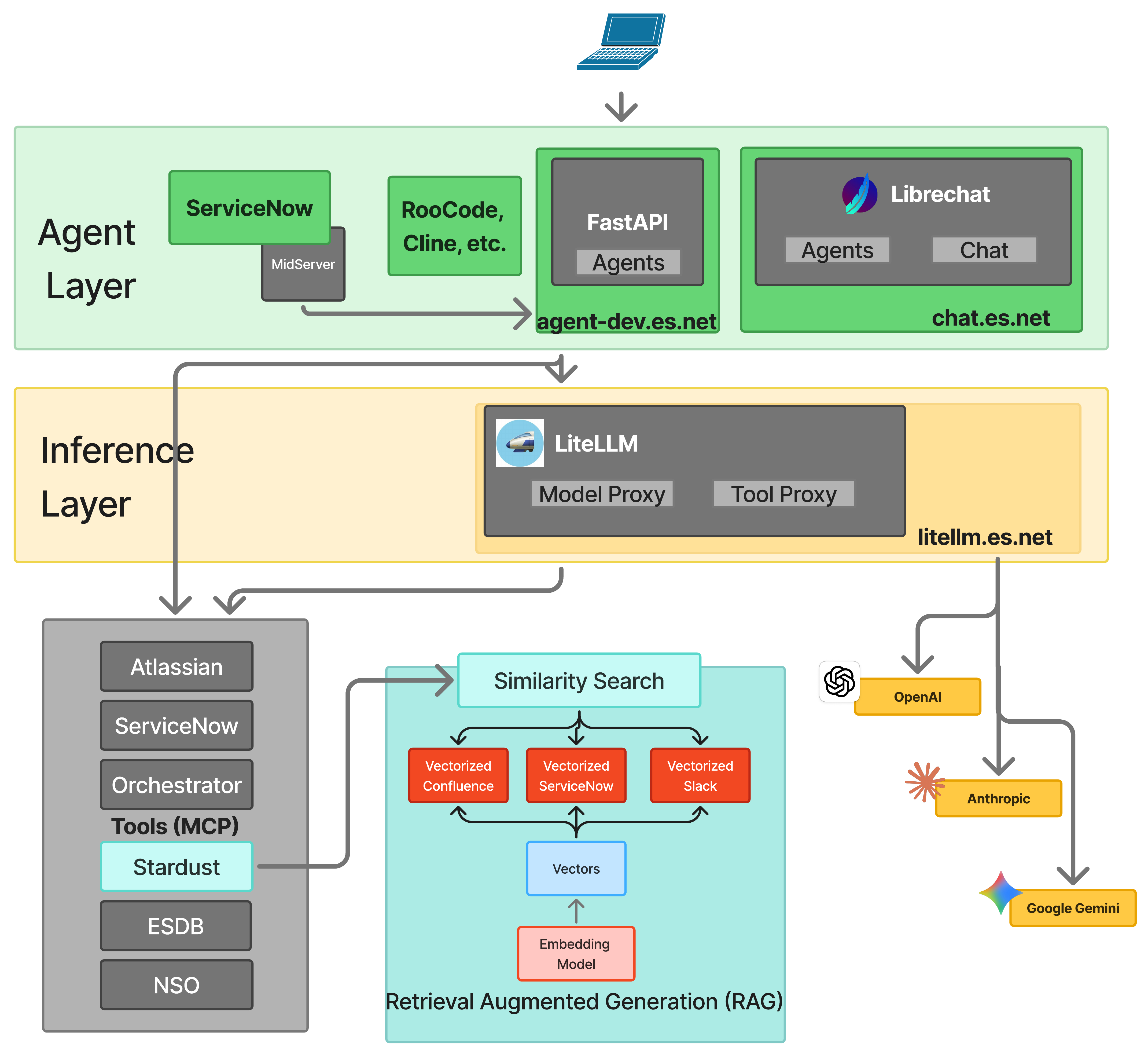}}
  \caption{Detailed view of the ORBIT system, with data access functions expanded
to show each MCP service available as of June 2026. The ServiceNow AI
Actions, invoked through the MID server, are the only ORBIT-specific
component; all other components are shared infrastructure designed to
support future use cases as well.}
  \label{fig:system-view}
\end{figure*}

ORBIT is implemented as an agentic, workflow-integrated system built to
support operator-led incident handling from within ServiceNow. The
overall architecture follows a layered design that separates orchestration,
retrieval, generation, and integration with existing operational tools
(see~\autoref{fig:logical-view} for a logical view and
\autoref{fig:system-view} for an implementation view). This separation is
necessary to achieve two properties simultaneously: (i) the ability to
ground AI outputs in operational evidence drawn from multiple ESnet sources,
and (ii) operational reliability through controlled tool invocation and
human oversight.

\subsection{Layered components and data flow}
The implementation comprises six functional components, the first four of
which are likely to be used in other AI-related projects, while the last two
are more specific to the particular tasks listed in \autoref{tb:tasks}.
The following is a brief description of these six components.

\paragraph{ServiceNow MID server workflow integration} ServiceNow controls
the event loop by invoking the agent at appropriate times (e.g., on new
incident tasks or specific business-rule triggers). The MID server receives
agent outputs and writes them into designated ticket fields using discrete
beta fields during development phases to preserve operational safety. This
component ensures minimal disruption to existing incident response processes
and enables accurate instrumentation of agent execution outcomes, more
in~\ref{sec:snow}.

\paragraph{Chat interface} A locally hosted web interface (chat.es.net)
provides the environment for interactive development and operator-facing
experimentation. It is implemented with an open-source tool known as
LibreChat~\footnote{Available at \url{https://www.librechat.ai/}}. It
supports prompt iteration, MCP-based tool access, and user evaluation
workflows (see~\autoref{sec:chat}). With the required data sources,
this interface also serves as the solution to Task 6 in \autoref{tb:tasks}.

\paragraph{Model Accesses via LiteLLM} The architecture routes LLM
calls through an enterprise proxy that standardizes access, monitors token
and other resource usage, and supports cost accounting. This design ensures
that usability and performance evaluation can include operational cost
signals, not just generation quality (see~\autoref{sec:litellm}).

\paragraph{Agentic application service} The core reasoning and
composition logic resides in a dedicated agentic application service,
accessed by ServiceNow AI Actions (and other clients) through a
REST-based API. Given a request type (e.g., ``resolution guidance''
or ``incident summary''), the agent orchestrates a sequence of tool
calls: querying MCP endpoints for ticket context, retrieving evidence
from the semantic search layer, optionally incorporating additional
contextual entities such as configuration item (CI) identifiers, and
performing LLM-based synthesis constrained by the retrieved evidence.
Prompt templates are versioned alongside the agent code and selectable
per request, so that prompts are managed as engineered, testable
artifacts rather than \emph{ad hoc} experiments (see Lesson~1 in
Section~\ref{sec:lessons}). Further architectural details are
described in~\autoref{sec:orb-srv}.

\subsection{Orchestration and event loops}
The key operational principle is a controlled event loop: ServiceNow
requests information from the agent as tickets are created or updated, and
the agent responds by performing retrieval and generation steps under
explicit tool constraints. This approach keeps the NOC workflow stable --
operators see AI outputs as structured additions within the incident record
rather than relying on a separate decision-making system. It also supports
reproducibility: each AI response corresponds to a request type, ticket
context snapshot, and evidence retrieval trace.

\paragraph{MCP tools for accessing ESnet data sources} ORBIT's agentic
layer reaches operational data through a set of Model Context Protocol (MCP)
tools, each exposing a narrowly scoped interface to a specific ESnet data
source. A \emph{ServiceNow MCP tool} provides read access to incident
narratives, work notes, and alarm/context fields required by each use case,
as well as controlled write-back pathways through predefined ServiceNow
mechanisms rather than \emph{ad hoc} ticket modifications. An \emph{ESDB MCP tool}
returns equipment and configuration information for network devices
referenced in an incident. There are a small number of MCP tools for
various other networking functionalities, not called out in
\autoref{tb:tasks}.

Complementing these structured-data interfaces, a unified
\emph{enterprise semantic search} layer vectorizes and indexes textual
knowledge sources, including Confluence procedures, ServiceNow
operational records, and Slack discussions, to support both semantic
similarity and hybrid search strategies that improve context matching
beyond keyword retrieval. By combining dense and sparse embeddings with
lexical retrieval and reciprocal rank fusion, the search layer handles
query styles ranging from free-text procedural questions to exact
identifier lookups. The current deployment shares infrastructure with
the Stardust network telemetry system~\cite{Balas:2022:Stardust} but
indexes only textual sources; incorporation of structured network
telemetry into the retrieval layer remains future work. The search
architecture and query pipeline are detailed in~\autoref{sec:search}.

\subsection{Safety and Security}\label{sec:safety}
    Security is an active area of development for ORBIT, with ongoing work toward alignment with ESnet's institutional AI guidance for Large Language Models operating within enterprise environments. Given the access-control and audit requirements that come with those environments, security controls fall into two categories: those already implemented in the current ORBIT Server and those planned for future implementation.

    \paragraph{Implemented controls} The current ORBIT Server provides immutable logging of all system interactions, tool configurations, and invocation arguments, providing a full audit record suitable for forensic review and ongoing performance monitoring. It also includes prompt injection mitigations, such as input validation and output filtering, to prevent sensitive material from appearing in model responses.

    \paragraph{Pending implementation} Service accounts are pending scoping to a predefined set of validated actions rather than broad operational permissions, with write scope to be restricted to prevent \emph{ad hoc} or unauthorized modifications to system state. Data retrieval will similarly be bound to what is relevant to the requesting user's context, preventing privileged information from surfacing where it does not belong.

\subsection{Governance, and evaluation}
Several design elements support evaluation and governance. First, the
system is human-in-the-loop by construction: it provides suggestions and
draft text, while operators remain responsible for acceptance and final
decision-making. Second, it includes feedback mechanisms in the ServiceNow
workflow so that operator judgments of usefulness and relevance can be
collected without separate tooling. Third, it logs LLM calls and exports
usage/performance telemetry using OpenTelemetry, enabling time-series
analysis of system behavior and cost signals over the research period.
These logs support deployment decisions by characterizing not just
whether the system runs, but how reliably it executes tasks, what
resources it consumes, and how operators interact with its outputs.

Finally, security requirements shape integration choices. Communications
between components occur over HTTPS with ESnet-approved authentication
methods, and ESDB access is protected via filtering in MCP interfaces and
controlled network access patterns during development. Write operations
occur only through the ServiceNow MID server's secured workflows with
attribution, auditability, and predefined action sets. Together, these
constraints support safe operational deployment while preserving the ability
to instrument the system for rigorous usability and performance measurement.

\subsection{Implementation Flexibility}
To support exploratory work within the fixed project duration while enabling
incremental learning, ORBIT follows a phased development model. The team
first iterates on query/prompt development and evaluation, then integrates
the resulting capabilities into the agent workflow and performs system-level
testing. Across phases, we refine the prompts, retrieval patterns, and tool
orchestration based on both operator feedback and measured execution
behavior, so that evaluation results can be rapidly translated into prompt
optimization and near-term decisions.

As an exploratory project, the team also remains explicitly flexible in
scope and implementation approach. In particular, as the underlying data
integration layer becomes available and more stable, we will be better
positioned to reassess the effort required to realize the task set in
\autoref{tb:tasks}. This flexibility allows us to pivot within the
allocation—prioritizing the tasks that show the strongest usability and
performance evidence, while potentially incorporating additional high-impact
tasks that can be implemented without extensive development time.

\section{Main Components} \label{sec:detail}
Following the high-level overview provided in the previous section,
this section provides more details on the six key components of the ORBIT
system: (1) the ServiceNow client, (2) the ORBIT agentic service, (3) the chat
interface, (4) semantic search, (5) the MCP tools for direct data access, and
(6) LiteLLM for model routing and accounting.

\subsection{ServiceNow Client through Agentic API} \label{sec:snow}

\begin{figure}
  \centerline{\includegraphics[width=0.48\textwidth]{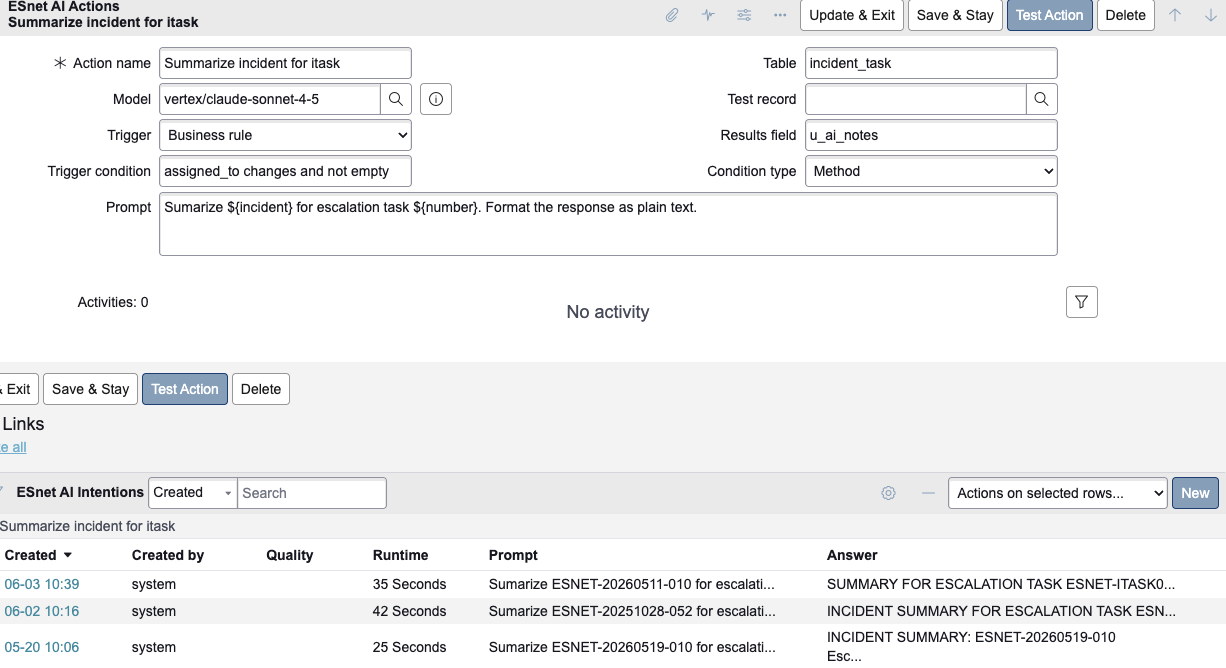}}
  \caption{A sample AI Action Form.}
  \label{fig:ActionForm}
\end{figure}

A central design principle of ORBIT is that AI assistance must meet
operators where they already work---inside ServiceNow---rather than
introducing another tool. Our primary design goal with the
ServiceNow-ORBIT integration was therefore to provide a transparent
interface for AI action tooling, including prompt design and trigger
points~\footnote{For developer documentation, see~\url{https://www.servicenow.com/docs/r/build-workflows/workflow-studio/flow-triggers.html}.}
throughout the ServiceNow platform (see~\autoref{fig:ActionForm}). Users
with a specific role can create, test, modify, and evaluate AI actions in
real-time with real data. ServiceNow AI actions are context-aware, which
enables field values to be inserted into each prompt to reference the
specific ticket or data element the prompt is running against. Once crafted
and optimized, AI actions are integrated into the system through one of
three mechanisms:

\begin{itemize}
    \item \textbf{Business rules} that fire automatically on defined conditions (e.g., creation of a new incident task), requiring no operator action;
    \item \textbf{Scheduled jobs} that run at predetermined intervals; or
    \item \textbf{UI actions} that can be triggered on demand by a NOC operator.
\end{itemize}

When an AI action is triggered on a specific record, a prompt is composed
from the prompt template associated with the action, resolving any field
variables it contains, and is passed to ORBIT via a REST-based API
through a ServiceNow MID Server. The current templates are listed
in~\autoref{tb:prompts}. The result is then stored in a field defined in
the AI Action and appears directly in the ticket's activity log, visible to
the entire team. Responses are logged as \emph{AI Intentions}, providing a
body of knowledge that can be reviewed for accuracy and ranked (\emph{Good},
\emph{Meh}, \emph{Bad}) with reviewer comments. This built-in feedback
mechanism supports systematic quality tracking, enabling the team to measure
output quality over time and identify where prompts or model selection
require refinement.

\paragraph{Model selection} Each AI action can be configured to invoke a
specific model drawn from a table of available models, all routed through
the LiteLLM gateway (Section~\ref{sec:litellm}). The current default for
most production actions is Claude Opus. Because model choice is
configurable per action, the team is collecting performance statistics to
inform cost-capability trade-offs, for example, assigning a lighter-weight
model to straightforward summarization tasks while reserving more capable
models for nuanced resolution guidance.

\paragraph{Operator control} Throughout this integration, the operator
always remains in control. AI-generated content is presented as a
suggestion that operators can accept, edit, or reject; the system functions
as an assistant, not a decision-maker. This human-in-the-loop design
preserves accountability and supports the governance requirements described
in Section~\ref{sec:safety}.

\paragraph{Avoiding response pollution} Originally, AI action results were
written to the \texttt{work\_notes} field of incident records. Because work
notes are a key input to actions such as ``Summarize Incident,'' a previous
AI response would be included in a subsequent invocation, polluting the
output with its own prior generations. As a solution, we created a custom
\emph{AI Notes} journal field to store AI action results. This resolved the
response pollution problem and also helped operators distinguish
AI-generated content from human-authored entries in the ticket activity
log---a separation that is equally important for preserving the integrity of
evaluation measurements (Section~\ref{sec:perf}).

\begin{table*}
  \caption{Prompt templates currently in use. The first five are for the
    tasks specified in~\autoref{tb:tasks}. The remaining ones were developed
    during software development (c.f.~\autoref{sec:changeAnalysis}). In
    these templates, the variable \$\{number\} will be substituted by the
    ticket/incident number and \$\{incident\} refers to the parent incident
    from which the new task is derived.}
  \label{tb:prompts}
  \centering
  \begin{tabular}{|p{1.0in}p{5.5in}|} \hline
\textbf{Action name} & \textbf{Prompt Template} \\ \hline
Summarize \newline incident \newline for hand-off &
Summarize \$\{incident\} for escalation task
\$\{number\}.  Format the response as plain text.
\\ \hline
Generate ticket \newline timeline &
For \$\{number\} in ServiceNow, generate a timeline of the ticket with all actions taken and resolution information for inclusion in an after action report.
\\ \hline
Suggest alarm \newline procedure &
Search the wiki docs for alarm procedure that are relevant for \$\{number\}.  Also review the history and emails in the ticket to see what has already been done.  Suggest the next step.  Be brief.  Format the results as plain text.
\\ \hline
Recommend \newline incident priority &
For \$\{number\} in ServiceNow.  Also review the documentation in the wiki about ticket and alarm priorities.  Recommend if any changes are needed to the priority of the incident or if the current priority is correct.  Respond with a one sentance answer.
\\ \hline
Propose \newline Final RFO &
For incident \$\{number\} in ServiceNow, propose the required fields for resolving the Incident.  Consult the appropriate Confluence wiki documentation in making the determination.  Specifically, state a proposed value for RFO, Resolved By, and a brief one or two sentence summary for Resolution Notes.  Respond in plain text with no markdown formatting.
\\ \hline \hline
Caused-By-Change \newline Analysis &
Look up the Incident \$\{number\}.  Also look up the CHG which is referenced in the Caused by Change field (which is named 'caused\_by' in the API).  If that field is blank, just report that and be done.  Otherwise, list if the Incident was created during the planned start and end times of the CHG and report that.  Include the timezones.  Next, check to see if the Configuration Item of the Incident is included in the list of Affected CI's for the CHG.  Use sys\_id's for that analysis.  If you find that the Incident was opened during the window and the Incident CI was in the Affected CI list of the Change then double check the results.  Report on that as well.  In the outputs, never include the raw 'sys\_id' fields.  Always use a short name or description or something meaningful to a human.  Format the response in plain text.  Do not use markdown.
\\ \hline
Change \newline Analysis &
Look up the "Incidents Caused by Change" related list of incidents.  list if each of the Incidents was created during the planned start and end times of the CHG and report that.  Include the timezones.  Next, check to see if the Configuration Item of the Incident is included in the list of Affected CI's for the CHG.  Report on that as well.  In the outputs, never include the raw 'sys\_id' fields.  Always use a short name or description or something meaningful to a human.  Format the response in plain text.  Do not use markdown.
\\ \hline
  \end{tabular}
\end{table*}

\subsection{ORBIT Agentic Service} \label{sec:orb-srv}

\begin{figure*}
  \centering
  \includegraphics[width=0.7\textwidth]{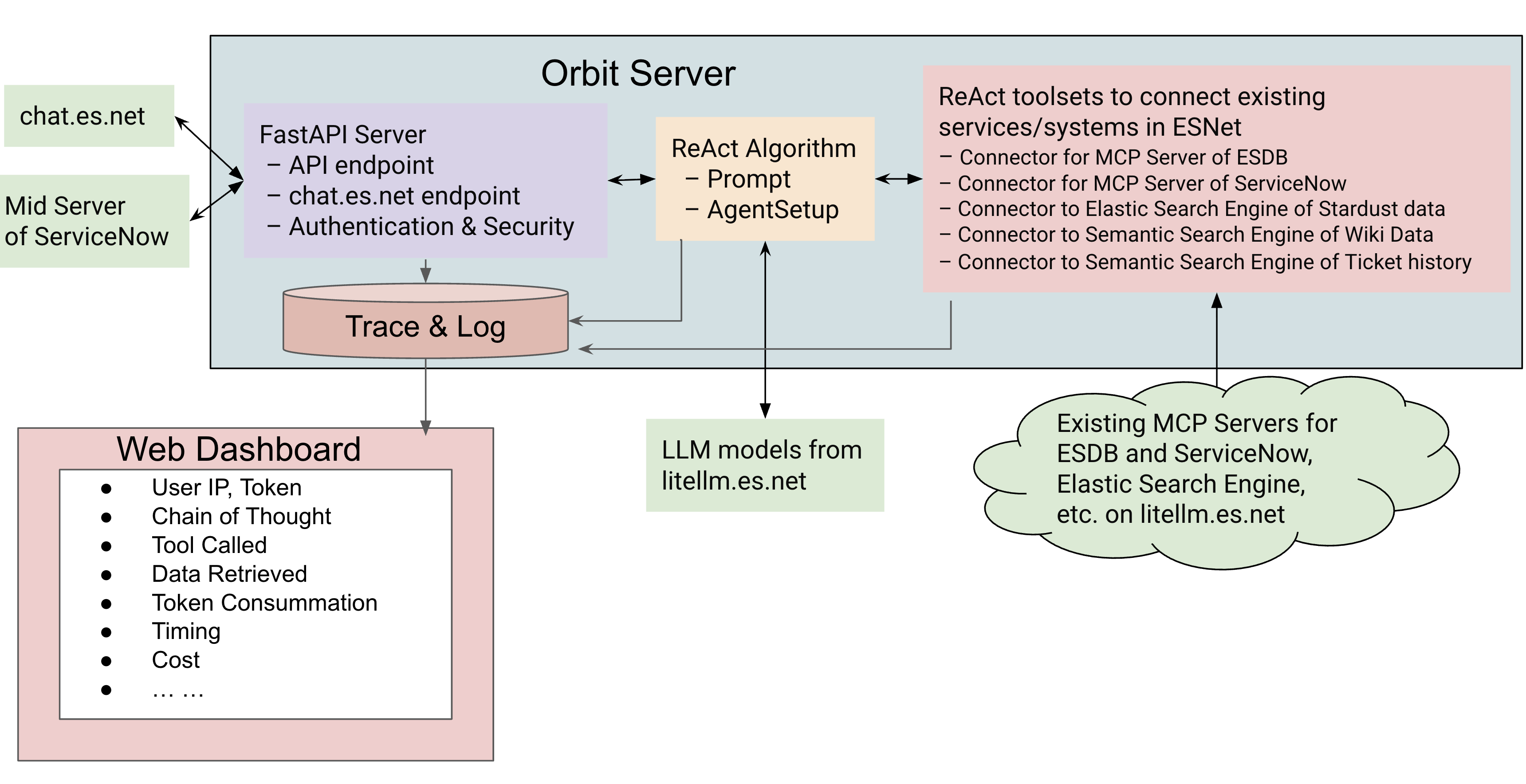}
  \caption{Overview of the ORBIT application service and its web
    dashboard. Through a series of connectors, the service interacts with
    MCP tools for data access and with AI models for reasoning and planning
    based on input prompts.}
  \label{fig:orb-srv}
\end{figure*}

\begin{figure*}
  \centering
  \begin{subfigure}[b]{\textwidth}
    \centering
    \includegraphics[scale=0.25]{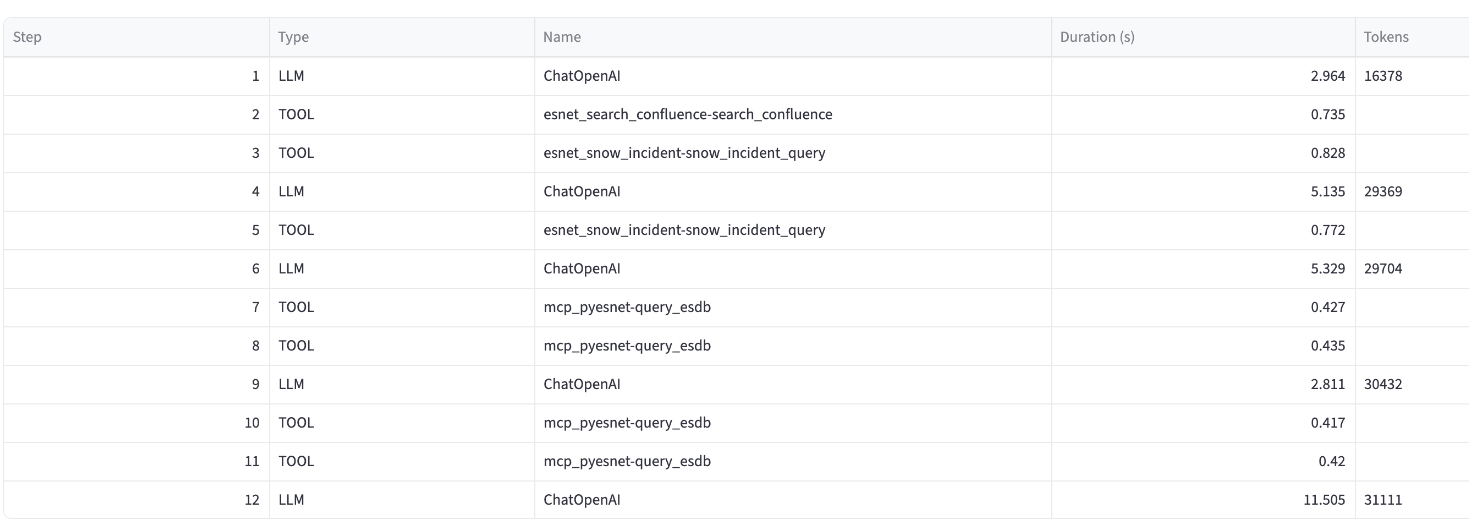}
    \caption{An example agentic action trace with 12 steps, including
      per-step time measurement and token usage.}
    \label{fig:trace-table}
  \end{subfigure}

  \vspace{1em}

  \begin{subfigure}[b]{\textwidth}
    \centering
    \includegraphics[scale=0.32]{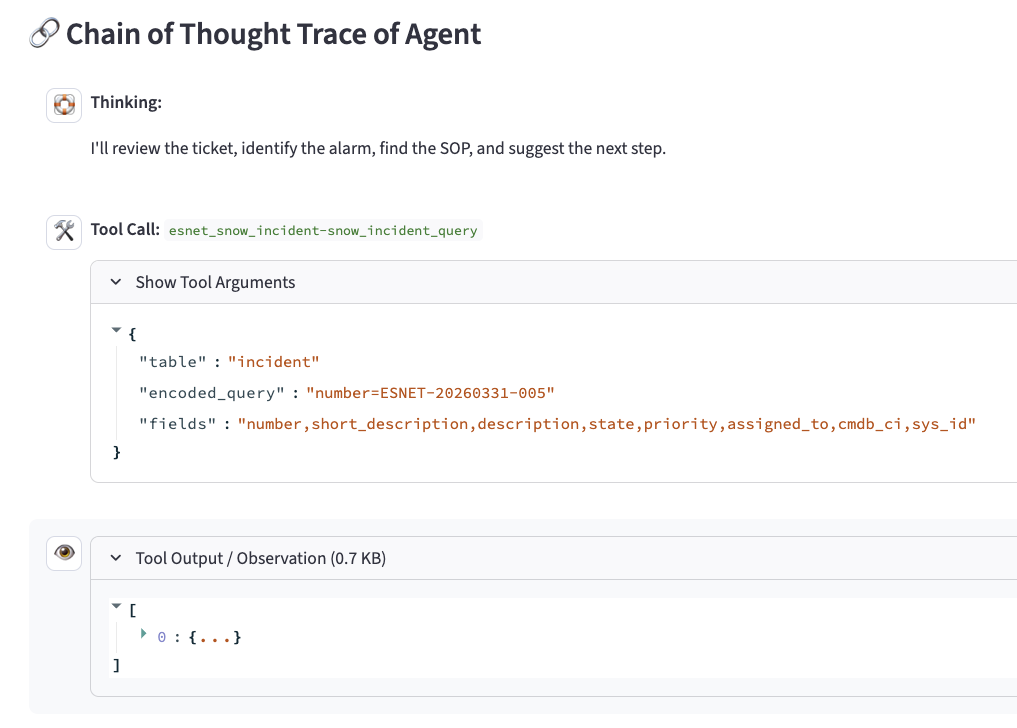}
    \caption{Detail of a single reasoning step, including the chain of
      thought, tool call with parameters, and tool output (omitted for
      brevity).}
    \label{fig:trace-step}
  \end{subfigure}

  \caption{A sample execution trace from the web dashboard for an instance
    of the task 3 ``Suggest alert procedure'' for a ticket.}
  \label{fig:dashboard}
\end{figure*}

The ORBIT agentic application service functions as the system's centralized
reasoning hub, responsible for integrating data from multiple operational
sources, orchestrating agentic workflows, and delivering AI-generated
outputs to downstream consumers. Its core responsibilities include
querying MCP endpoints for ticket context and equipment metadata,
performing semantic searches across vector databases, managing
interactions with large language models through LiteLLM, and synthesizing
the results into actionable summaries and resolution guidance via
Retrieval-Augmented Generation.

The service is designed for extensibility: its architecture supports the
incremental addition of new data connectors and query services, enabling
broader ESnet operational tasks beyond the initial ORBIT scope. Because
the four elements required for each execution---tools, prompt template,
language model, and agent executor---are all selected per request rather
than compiled into the service, the same server functions as a
general-purpose agentic composition service. New workflows can be
introduced by registering additional MCP tools and prompt templates
without modifying the service itself. By hosting this infrastructure
within ESnet's established environment, the system maintains strict
compliance with security policies and access controls for sensitive data.
A dedicated server also abstracts integration complexity from client
applications, allowing ServiceNow AI Actions, the chat interface, and
future consumers to invoke the same reasoning capabilities through a
uniform API.

\subsubsection{External Interactions}
As illustrated in~\autoref{fig:orb-srv}, the ORBIT application service
comprises three internally developed components, interfaces with three
existing ESnet infrastructure services, and connects to five external MCP
or Elasticsearch endpoints.

On the \emph{client-facing} side, the service provides the backend for
\texttt{chat.es.net}, enabling interactive ticket resolution, and receives
API calls from the ServiceNow MID server via a Python client or direct
POST requests. Because the MID server is pre-existing infrastructure,
integration required coordinating with the ESnet ServiceNow team to
implement targeted feature extensions rather than deploying new middleware.

On the \emph{data-facing} side, the service gathers operational context
through a set of specialized MCP connectors and search interfaces: a
ServiceNow MCP server for reading ticket narratives, work notes, and
alarm fields; an ESDB MCP server for equipment and configuration
metadata; Stardust MCP endpoints for router and interface telemetry; and
dedicated semantic search engines backed by Elasticsearch for retrieving
correlated Wiki pages and historically similar tickets. The reasoning
agent autonomously selects which of these tools to invoke at each step.
To reduce invocation overhead, the architecture explores importing MCP
tool definitions directly into the reasoning loop via specialized
adapters, eliminating the need for intermediary adapter servers on the
host machine. All LLM calls are routed through the LiteLLM gateway
(Section~\ref{sec:litellm}).

\subsubsection{Internal Components}

\paragraph{FastAPI server and request handling}
The service is built on FastAPI~\cite{fastapi}, which handles client
requests, invokes the core reasoning algorithm, and enforces security
parameters. It exposes distinct endpoints for OpenAI-compatible chat
requests and MID server API requests. Payloads encapsulate user inputs
and vary with dynamic system factors---for example, whether the server
fetches ticket information dynamically from a supplied identifier or
expects the client to provide full ticket context upfront, and whether the
end user may select a specific language model. The FastAPI server routes
processed inputs to an agent executor, which returns a final response
formatted as a plain-text string for chat interfaces or as a structured
JSON payload for the MID server.

\paragraph{Security and authentication}
Access control follows the principle of least privilege. At the network
level, the server is deployed within a protected ESnet segment accessible
only through the corporate VPN. At the application level,
machine-to-machine authentication uses static, token-based bearer
credentials; authorized clients receive unique, long-lived secret tokens
stored in approved secret managers rather than in source code.

For authorization, the server associates each request with the initiating
end user and forwards credentials as JSON Web Tokens. Downstream
services perform independent authentication and authorization, returning
standard HTTP~403 responses for unauthorized actions. All communication
between clients and the server is encrypted using Transport Layer Security
in accordance with ESnet policy.

\subsubsection{Reasoning Algorithm and Tooling}

The reasoning engine implements the ReAct algorithm~\cite{yao2023react},
which interleaves reasoning traces with action steps to iteratively
resolve ticket queries. Built on LangChain~\cite{langgraph2024}, the
architecture is model-agnostic, though it retains the flexibility to adopt
model-specific optimizations should a highly specialized model be
warranted. Each execution requires four elements: a suite of operational
tools, a versioned prompt template tailored to the task, a designated
language model, and an agent executor. The system prompt governs the
model's operational logic and tool-selection strategy and is maintained
separately from the raw user inputs received through the chat interface.

To support concurrent users, the FastAPI server combines an asynchronous
event loop with a stateless agent design: each incoming request
instantiates an independent agent with no persistent session history.
While session-based or agent-pool architectures could offer persistent
memory or improved resource utilization, the stateless approach ensures
clean, isolated execution for every request in the production environment.

\subsubsection{Execution Traces and Continuous Monitoring}

Every tool invocation and reasoning step is recorded by the service's
\emph{web dashboard} (Figure~\ref{fig:dashboard}), which provides a
detailed execution trace for each agent run. For every request, the
dashboard captures the sequence of tools called, the data accessed, token
consumption, step-level timing, and an estimated execution cost. This
instrumentation serves two purposes: it gives developers the information
needed to analyze and debug individual agent runs, and it surfaces
system-wide patterns, such as frequently invoked tools, latency
bottlenecks, and cost outliers, that guide prompt optimization and
resource planning.

Beyond \emph{ad hoc} debugging, the traces also underpin a continuous monitoring
regime. Drawing on feedback from NOC engineers across more than
200~instances of ServiceNow AI actions, the team distilled a regression
suite of approximately four dozen tests. These tests are executed daily
against production-representative tickets, providing longitudinal
performance metrics that detect prompt regressions, tool-selection
drift, and latency changes before they affect operators.

\subsection{Chat Interface} \label{sec:chat}

Our team has also implemented a chat interface hosted on \texttt{chat.es.net},
a conversational AI platform built on the open-source LibreChat project.
Within the context of ORBIT, its primary contribution has been to provide an
accessible, interactive interface for experimentation: it lets users iterate
on prompts, compare model behavior, and validate agent and tool integrations
without writing code or standing up their own client. This lowers the
barrier to participation, allowing both engineers and non-developers to
explore how large language models respond to operational network data and to
refine the prompting strategies that downstream ORBIT components ultimately
depend on.

The platform brokers model access through the LiteLLM proxy
(Section~\ref{sec:litellm}) and exposes ESnet data sources through MCP server
integrations (Section~\ref{sec:mcptools}). Notably, because access is gated by
ESnet's single sign-on, \texttt{chat.es.net} is currently the one path in which
tool invocations can be bound to an authenticated end-user identity rather than
a shared service credential---a distinction we return to in
Section~\ref{sec:mcptools}. This makes the platform a natural staging ground for
testing tool integrations before they are promoted into the more tightly scoped,
production-facing agents described in Section~\ref{sec:orb-srv}.

More broadly, \texttt{chat.es.net} serves as ESnet's general-purpose AI
workbench, giving staff a sanctioned, secure environment for everyday
tasks---summarization, drafting, code assistance, and data exploration---while
keeping interactions inside ESnet's trust boundary rather than on external
commercial services. It therefore plays a dual role: a productivity tool for the
wider organization, and a low-risk proving ground where the prompts, agents, and
integrations that support ORBIT are developed and refined.

\subsection{Semantic Search} \label{sec:search}

\begin{figure}
  \centerline{\includegraphics[width=0.48\textwidth]{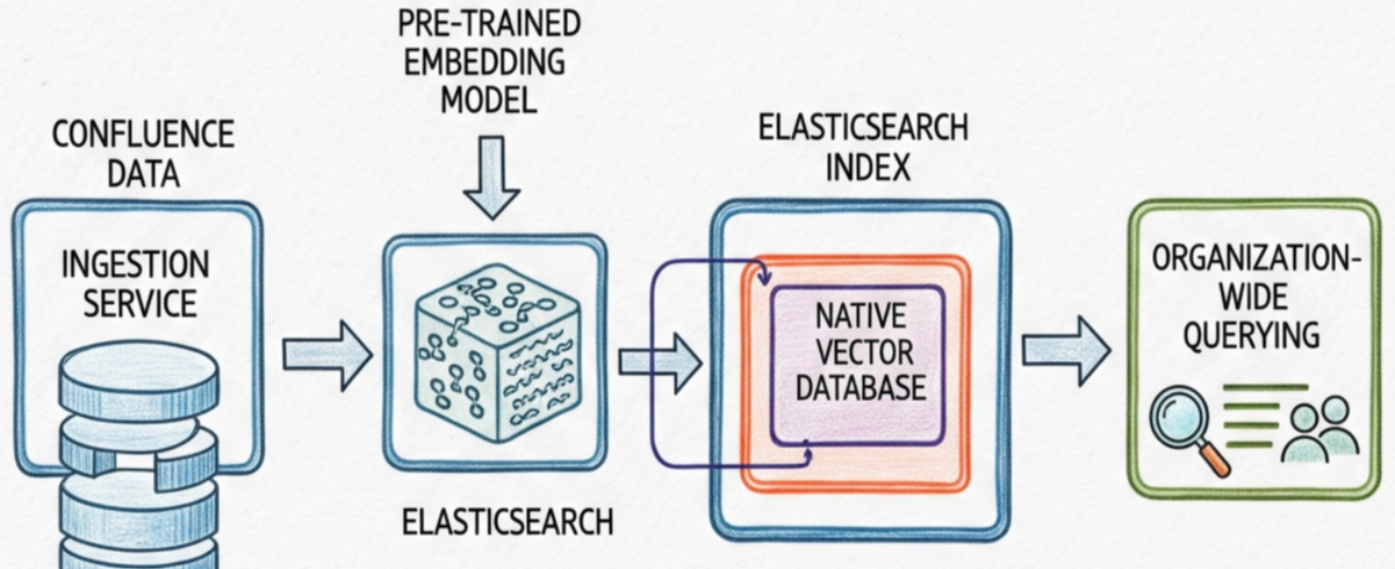}}
  \caption{A high-level illustration of the semantic search pipeline in the Elasticsearch platform.}
  \label{fig:search}
\end{figure}

ESnet's Hybrid Semantic Search platform unifies operational knowledge currently scattered across ServiceNow, Confluence (wiki), and Slack into a single, AI-powered retrieval layer. The system is built on Elasticsearch and served through existing interfaces (e.g., LibreChat and MCP servers), enabling users to issue natural-language queries without manually selecting a target system (see~\autoref{fig:search}).

\subsubsection{Indexing pipeline}
Data is ingested into source-dedicated Elasticsearch indices to preserve source-specific structure and allow per-source tuning.

\begin{itemize}
\item  ServiceNow content is exported via a custom script that captures resolved/closed incidents and key fields (e.g., short description, narrative, resolution notes).
\item Confluence and Slack are synchronized using Elastic’s native connectors.
\end{itemize}

For semantic retrieval, the platform generates embeddings using carefully selected fields rather than embedding all text. Field selection is critical for precision: ServiceNow emphasizes incident descriptions and resolution text, while Confluence emphasizes page titles and bodies, and Slack emphasizes message text and thread context.

The embedding stage runs two complementary models:
\begin{enumerate}
\item ELSER (Elastic Learned Sparse EncodeR) produces sparse representations by expanding text into weighted token distributions over a high-dimensional vocabulary, capturing learned associations that enable retrieval across vocabulary mismatch.
\item E5 (Multilingual-E5-Small) produces dense vector embeddings that encode semantic meaning, enabling similarity search based on conceptual proximity rather than term overlap.
\end{enumerate}
These embeddings are indexed to support both sparse semantic and dense semantic retrieval, complementing the lexical retrieval provided by BM25.

\subsubsection{Hybrid querying architecture}
The system issues a single query containing four sub-retrievers, each optimized for different query characteristics. Elasticsearch executes these sub-retrievers internally and merges their ranked lists via RRF before returning results:
\begin{enumerate}
\item ELSER sparse retrieval uses a learned sparse encoder to expand both documents and queries into weighted token distributions over a high-dimensional vocabulary. These expansions capture learned associations rather than synonyms, enabling retrieval of semantically relevant incidents even when no query terms appear verbatim in historical records — without requiring explicit synonym configuration.
\item E5 dense retrieval performs approximate nearest-neighbor (ANN) search for semantic similarity, even when no relevant terms overlap. Vectors are indexed using BBQ-HNSW, which combines 1-bit quantized candidate retrieval with automatic oversampled rescoring against full-precision vectors to preserve recall under compression.
\item BM25 with field boosting provides lexical precision for exact technical terminology, with boosts reflecting that certain fields (e.g., incident summaries or root-cause fields) are more semantically informative.
\item Exact identifier matching is a lightweight keyword-based retriever for structured identifiers (e.g., incident/ticket IDs, circuit IDs, CI names), implemented as exact term queries on keyword fields. Because term queries return no results when the query does not match a stored identifier exactly, this component is naturally inert for free-text queries and contributes only when structurally appropriate.
\end{enumerate}

\subsubsection{Fusion and final ranking}
The four ranked lists are merged using Reciprocal Rank Fusion (RRF). RRF combines rankings using relative positions rather than raw score magnitudes, avoiding score-scale incompatibilities across retrievers. The result is a single consensus ranking that remains robust across diverse operational query styles---from free-text questions to structured identifier lookups.

\subsection{Tools for Direct Data Accesses} \label{sec:mcptools}
ORBIT's reasoning agent accesses operational data through a set of
purpose-specific tools, each scoped to a well-defined data domain rather
than exposed as a general-purpose query interface. The current tool set
comprises:

\begin{itemize}
  \item A \emph{ServiceNow tool} that retrieves incident payloads---including
    narratives, work notes, and alarm/context fields---for a given ticket
    identifier;
  \item An \emph{ESDB tool} that returns equipment and configuration
    metadata for network devices referenced in an incident;
  \item \emph{Semantic search tools} that retrieve correlated Confluence
    Wiki pages and historically similar tickets
    (Section~\ref{sec:search}).
\end{itemize}

Each tool is implemented as an MCP tool, ensuring that the agent
receives structured, validated responses while data owners retain
explicit control over which fields and operations are reachable.
Restricting each tool to a narrowly defined operation---rather than
exposing a broad query surface---is a deliberate security decision: it
minimizes the attack surface and eliminates the need for fragile input
validation on free-form queries.

Because these tools were developed incrementally alongside existing ESnet
systems, they are reached through three distinct access mechanisms, each
selected to match the requirements of a particular integration context:

\begin{itemize}
  \item \textbf{In-process tools} are imported directly into the agent's
    reasoning loop (Section~\ref{sec:orb-srv}), avoiding the overhead of
    an intermediary adapter and providing low-latency access for tightly
    coupled, high-frequency invocations.

  \item \textbf{Gateway-brokered tools} are accessed through the LiteLLM
    MCP gateway (Section~\ref{sec:litellm}), which centralizes access
    control, usage accounting, and observability across the broader AI
    ecosystem, making each tool reusable by applications beyond ORBIT.

  \item \textbf{Identity-gated tools} are surfaced through
    \texttt{chat.es.net} (Section~\ref{sec:chat}), which inherits the
    platform's OpenID Connect (OIDC) authentication and binds each tool
    invocation to an authenticated end user.
\end{itemize}

Each mechanism provides adequate security and functionality
within its own scope; collectively, they reflect the practical reality
that operational AI tooling must balance competing concerns---invocation
latency, centralized governance, and end-user identity
propagation---that no single integration pattern satisfactorily addresses
today. Identity propagation illustrates this tension: requests traversing
machine-to-machine boundaries carry a forwarded JSON Web Token that the
receiving service validates independently
(Section~\ref{sec:orb-srv}), whereas requests originating from an
interactive session arrive already bound to an authenticated user.
Converging these mechanisms behind a unified, identity-aware access layer
is an area of active interest both at ESnet and across the broader
community; we discuss this further in Section~\ref{sec:conclusion}.

\subsection{Model Access Gateway} \label{sec:litellm}

ORBIT operates within a centralized AI gateway ecosystem built around
LiteLLM, an open-source model proxy and routing layer that provides
unified access to large language models and AI tools across ESnet's
infrastructure~\cite{LiteLLM}. Rather than requiring each application to
manage its own provider credentials, model selection logic, and usage
accounting, the gateway consolidates these concerns behind a single
authenticated entry point. This allows for enforcing usage policies, enabling cost
accountability across users, projects, and organizational groups, and
ensuring consistent security governance. The key components of this
ecosystem are shown in~\autoref{fig:system-view}.

\paragraph{Gateway architecture}
The LiteLLM gateway is deployed as a GitOps-managed Kubernetes
application on ESnet infrastructure, with continuous delivery handled by
ArgoCD. It exposes an OpenAI-compatible API endpoint, providing broad
compatibility with existing AI tooling and developer workflows without
requiring custom client integrations.

Model traffic is routed across four provider groups:
\begin{itemize}
  \item Anthropic Claude models (Haiku, Sonnet, Opus) via Google Vertex~AI,
  \item Google Gemini models via Vertex~AI,
  \item third-party commercial and open-source models available through
    Vertex~AI, and
  \item on-premises models hosted by Lawrence Berkeley National
    Laboratory's CBORG service.
\end{itemize}
Commercial requests are routed through FedRAMP-compliant Google Cloud and
AWS infrastructure, providing data-handling protections consistent with
institutional requirements. On-premises CBORG models offer zero-cost
alternatives suitable for routine workloads, while cloud-hosted commercial
models serve more demanding or complex tasks. This tiered arrangement
allows the team to balance capability against cost on a per-task basis.

\paragraph{Authentication and access control}
The gateway issues virtual API keys scoped per user, application, or
project. Each key carries role-based access control (RBAC) policies that
govern which models, tools, and data resources it may reach.
Machine-to-machine workflows are supported through dedicated API keys,
enabling automated pipelines and agentic systems to authenticate without
user intervention. Fine-grained downstream key management allows ESnet to
retain control over upstream provider credentials while distributing
access broadly: when backend resources change, such as when new models are added,
tools are updated, or providers are rotated, end users and applications
require no re-keying, as the gateway handles routing transparently.

\paragraph{MCP tool integration}
Beyond model access, the gateway proxies MCP tools that expose internal
ESnet data sources, monitoring systems, documentation repositories, and
operational APIs to LLM-enabled applications. These tools are aggregated
into a unified namespace accessible through a single authenticated
endpoint. The gateway enforces tool-level authorization policies,
ensuring that each application can reach only the specific tools its API
key permits. This capability addresses a significant governance gap, as
many MCP tool implementations lack native authentication mechanisms.
Tool sets can further be filtered to the subset relevant to a given
application, reducing unnecessary context consumption during LLM
interactions.

\paragraph{Usage tracking and cost accounting}
The gateway records all model invocations and MCP tool calls, attributing
resource consumption to individual users, API keys, teams, and projects.
Usage data is persisted in a Patroni-managed PostgreSQL cluster,
providing durable records for reporting and auditing. Administrative
dashboards offer real-time visibility into spending patterns, model
utilization, and tool invocation rates, while configurable spending limits
and alerts enable project managers to enforce budgets proactively. This
accounting infrastructure is particularly important given the mixed cost
profile of the model portfolio: CBORG on-premises models carry no direct
cost, whereas commercial cloud models are billed per token and must be
tracked against institutional contracts and per-user spending limits.

\paragraph{Security and compliance}
The gateway serves as a centralized security enforcement point,
consolidating request and response logging, prompt-injection protection,
data-leak prevention, and a comprehensive audit trail for all model and
tool interactions. This unified visibility enables security teams to
detect anomalous usage patterns and enforce data-handling policies
consistently across all AI-enabled applications, both within ORBIT and
across the broader ESnet environment.

\section{Performance and Usability} \label{sec:perf}

This section reports the experimental results for ORBIT, an AI assistant
integrated into an NOC incident workflow to support six operator-selected
use cases, with emphasis on Intelligent Resolution Guidance and Adaptive
Incident Summarization. The evaluation is designed to address a core
concern in production AIOps: improvements must be demonstrated through
operational usability and measurable performance, not through model novelty
alone. Accordingly, our experiments quantify (i) how effectively ORBIT
reduces operator effort, (ii) how reliably it executes the intended workflow
actions under real operational constraints, and (iii) how prompt and
orchestration choices affect both response quality and system behavior.

Usability and adoption measurements are collected through structured user
evaluation with NOC operators. In addition to survey-based feedback (e.g.,
relevance/usefulness ratings), we also use observational studies to capture
how operators interact with ORBIT outputs in ServiceNow---particularly
whether the assistant helps operators find correct context faster, whether
it changes the operator's search-and-retrieval process, and whether outputs
are accepted, edited, or rejected. These measurements are paired with
interface-level signals that reflect real workflow engagement, such as
acceptance patterns and user feedback on generated suggestions.

Performance measurements focus on actionable system characteristics that are
observable during workflow execution. We track end-to-end responsiveness
for ORBIT actions, the number of underlying agent/tool steps executed per
request, and resource consumption proxies such as LLM token usage (and
related cost considerations). While global metrics like MTTR may be
difficult to attribute cleanly due to noisy incident data and limited sample
sizes, we instead measure targeted timing that directly reflects operator
work---most notably the time operators spend researching and validating
ticket context. We also instrument latency and retry behavior to
characterize the reliability of agent execution in the presence of
stochastic model outputs.

Finally, the experiments include a dedicated analysis of prompt
optimization. We treat prompts as executable artifacts (``prompts are
code'') and evaluate prompt variants using controlled incident replays.
Results show that engineered, stepwise prompts can reduce the number of
agent actions and improve first-try correctness, while careful orchestration
mitigates stochastic failures (e.g., incorrect tool selection) observed when
prompts and tool descriptions are ambiguous. Together, these measurements characterize ORBIT along dimensions that
matter for deployment decisions: not only whether outputs are plausible,
but whether they are consistently useful, grounded, and operationally
efficient enough to inform next-step deployment and broader AIOps
investment.

\subsection{Statistics from ServiceNow AI Actions} \label{sec:aiactions}

We begin our experimental measurements with the AI Actions from the
ServiceNow system for handling incidents reported to ESnet. The evaluation
period covered in this section runs from January through June~2026
(approximately six months). As of the writing of this report, the
ServiceNow system records a total of 169~AI Actions over that period. The
statistics reported here were gathered during the development and initial
deployment phase and largely reflect activity to date; further changes are
expected as prompt optimization and broader operator engagement continue.

\begin{table}
  \caption{The number of times ServiceNow AI Actions are invoked by
    different types of users.
    Note ``DEV'' for developers, ``NOC'' for NOC engineers, and ``SYS'' for
    automated actions triggered by ServiceNow rules.}
  \label{tb:acts-by-user}
  \centering
  \begin{tabular}{|l|r|} \hline
DEV &  41 \\ \hline
NOC &  12 \\ \hline
SYS & 116 \\ \hline
  \end{tabular}
\end{table}

In \autoref{sec:snow}, we provided a detailed description of how the
  ServiceNow AI Actions might be triggered, for example by an individual
  user through UI Actions or by systems through Business Rules.
  \autoref{tb:acts-by-user} shows that nearly 70\% of the total 169 AI
  Actions are triggered automatically (116 out of 169). Only about 7\%
  of AI Actions are triggered through UI Actions, and many users appear
  to have tried it only once. Nearly a quarter of the AI Actions are
triggered by developers, which is not surprising given that these
statistics were gathered during the development phase of the project.

\begin{table*}
  \caption{Statistics of ServiceNow AI Actions available as of June 2026.
    The numbered actions are those defined in~\autoref{tb:tasks}, while
    the un-numbered ones were recommended by NOC engineers after the start of
    the project.}
  \label{tb:acts-by-task}
  \centering
  \begin{tabular}{|ll|r|r|r|} \hline
 & \textbf{Task} & \textbf{Count} & \multicolumn{2}{c|}{\textbf{Duration} (s)} 
    \\ \hline
 & & & Average & StdDev \\ \hline
1 & Summarize incident & 29 & 30.4 & 16.3\\ \hline
2 & Generate time line & 6 & 61.3 & 48.0 \\ \hline
3 & Suggest alarm procedure & 55 & 50.5 & 59.0 \\ \hline
4 & Recommend alarm priority & 48 & 33.1 & 50.8 \\ \hline
5 & Propose Final RFO & 8 & 27.4 & 14.1 \\ \hline
\hline
& Caused-By-Change Analysis & 21 & 35.0 & 17.5 \\ \hline
& Change Analysis & 2 & 88.5 & 122.3 \\ \hline
  \end{tabular}
\end{table*}

Among the currently recorded AI Actions, \autoref{tb:acts-by-task} shows a
breakdown by defined tasks (see~\autoref{tb:tasks} and
\autoref{tb:prompts}). Note that the time reported here are recorded by
ServiceNow and more detailed time breakdown would be available from the
agentic application service, as described next in
\autoref{sec:server-stats}.

The most frequently invoked task is Task \#3 (suggest alarm procedure),
followed closely by Task \#4 (recommend alarm priority). Both are
recommendation tasks that require the ORBIT system to correlate information
from multiple sources and thus exhibit large variation in execution time.
From \autoref{tb:acts-by-task}, we see that their standard deviations
(StdDev) exceed their average execution times. In contrast, the
summarization tasks (Tasks \#1 and \#2) have smaller standard deviations
relative to their average execution time, which indicates that these tasks
involve more predictable data sources and run more consistently.

It is important to note that these invocation counts and timing statistics
do not capture output quality. As part of the evaluation, NOC engineers
have reviewed the AI-generated outputs for hundreds of incidents, assessing
whether the summaries, procedure suggestions, and priority recommendations
were accurate and operationally useful. That review process is ongoing,
and a more systematic quality evaluation---covering output accuracy,
operator acceptance rates, and task-by-task effectiveness---is planned for
the coming months as part of the project's next evaluation phase. Next, we
examine the components supporting these AI Actions in more detail.

\subsection{ORBIT Server Usage} \label{sec:server-stats}

\begin{figure}[htbp]
    \centering
    \includegraphics[width=0.9\columnwidth]{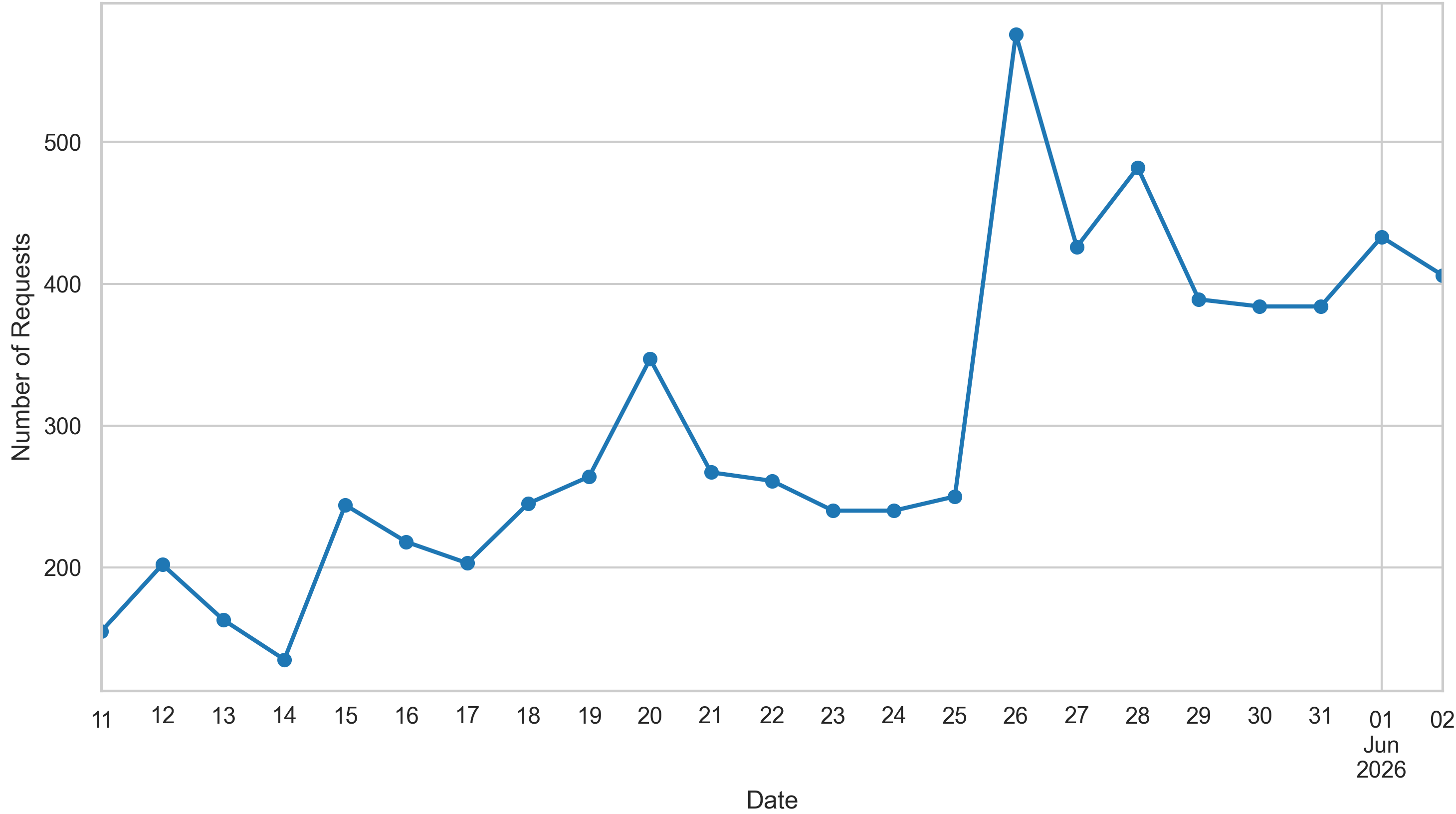}
    \caption{Daily total requests handled by the ORBIT Server from mid-May to June 2026. The generally increasing use is primarily due to the addition of new automated testing.}
    \label{fig:daily_requests}
\end{figure}

\autoref{fig:daily_requests} shows the overall utilization of the ORBIT server environment over time. It highlights peak usage periods, which generally correlate with active troubleshooting sessions or testing phases, giving a clear view of when the system experiences the highest load. Throughout May 2026, more regression test cases were developed based on NOC operator feedback. The total number of these test cases reached about four dozen by June. These test cases are executed on a daily schedule. Since the number of requests to the ORBIT Server stabilized around 400, which is much larger than four dozen, our observation is that there are more requests from other sources beyond ServiceNow AI Actions (see \autoref{sec:aiactions}), such as \texttt{chat.es.net} and Slack~\footnote{In collaboration with the Charles Shiflett of the Platform Engineering group, ORBIT is available as an AI Agent in Slack starting in mid-May 2026.}.

\begin{figure}[htbp]
    \centering
    \includegraphics[width=0.9\columnwidth]{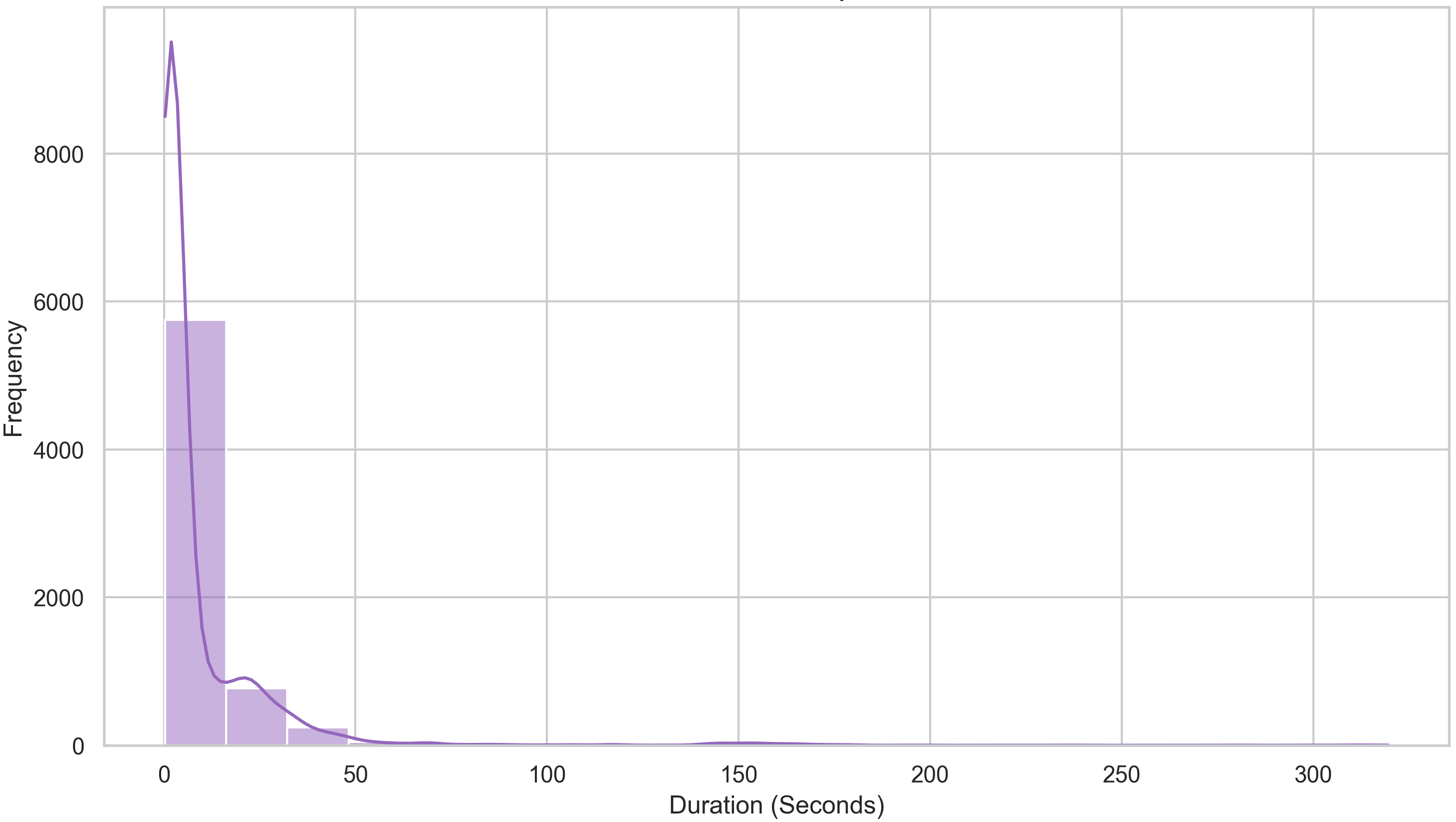}
    \caption{Distribution of total request durations on the ORBIT Server, highlighting typical execution latency and long-tail outliers.}
    \label{fig:duration_dist}
\end{figure}

Figure~\ref{fig:duration_dist} illustrates the spread of total request times from start to finish. Most agent executions cluster around a predictable median duration of a few seconds, while a noticeable long tail indicates complex queries requiring extensive tool use, retry logic, or prolonged language model generation time.

\begin{figure*}[htbp]
    \centering
    \includegraphics[width=0.75\textwidth]{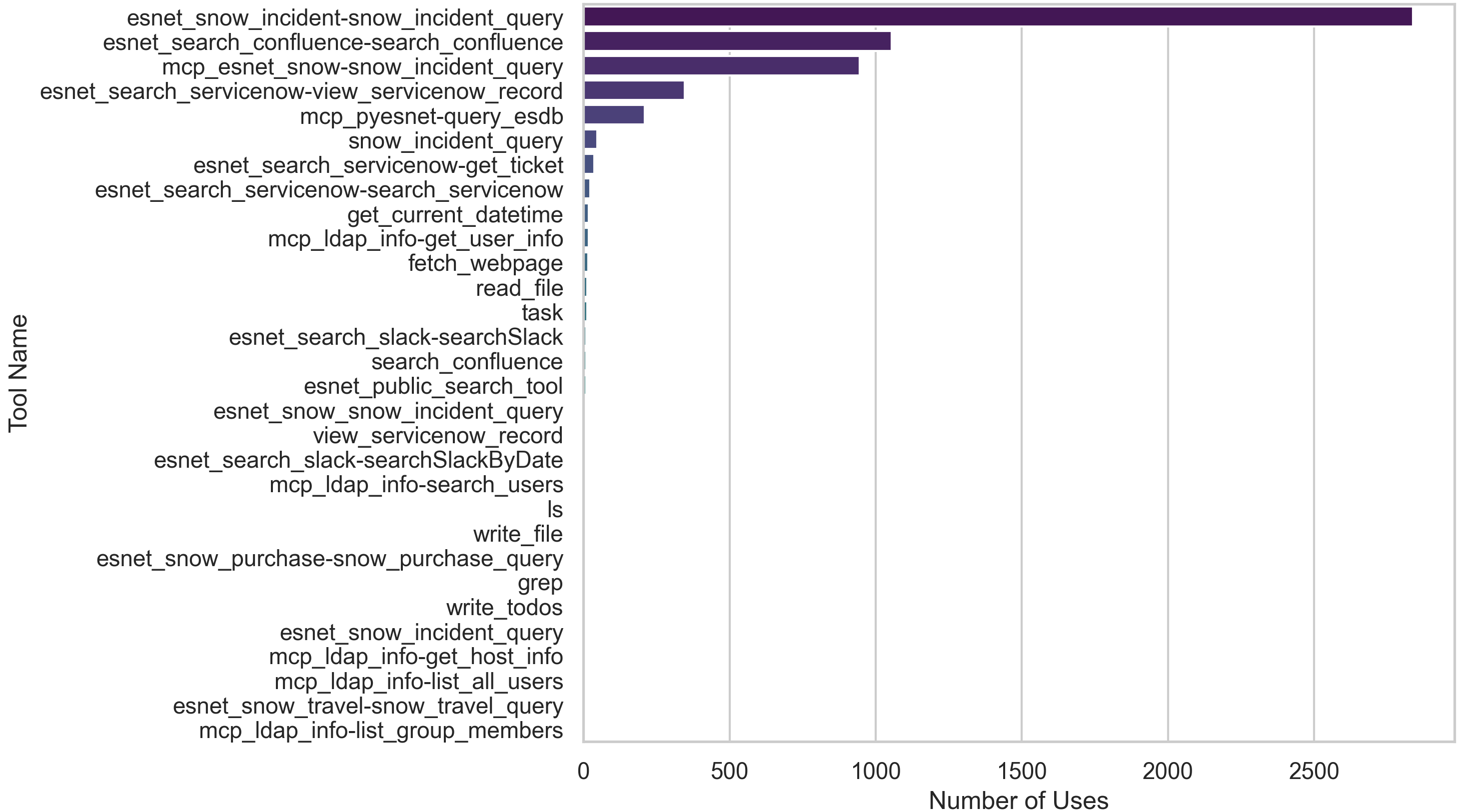}
    \caption{Most frequently used MCP tools on the ORBIT Server are for accessing incidents in ServiceNow and procedure documentation in Confluence.}
    \label{fig:tool_usage}
\end{figure*}

Figure~\ref{fig:tool_usage} reveals the agent's reliance on specific external systems to achieve its goals. The ServiceNow ticket retrieval tool and various Confluence standard operating procedure searches are the most frequently invoked actions, underscoring the agent's primary operational role in gathering incident context and cross-referencing internal documentation.

\begin{figure}[htbp]
    \centering
    \includegraphics[width=0.9\columnwidth]{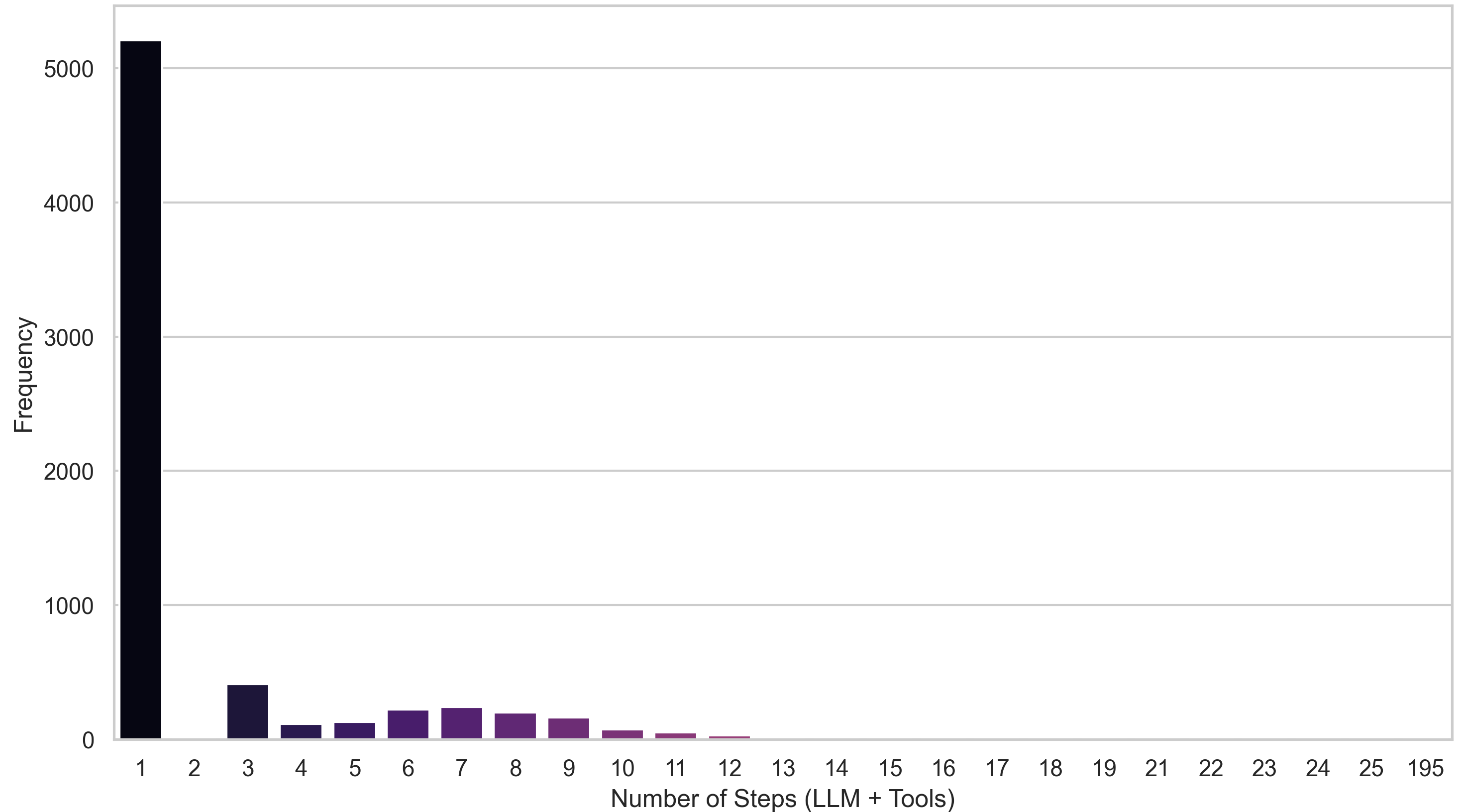}
    \caption{Distribution of execution steps per request on the ORBIT Server, representing the depth of the agent's iterative reasoning loop.}
    \label{fig:execution_steps}
\end{figure}

\autoref{fig:execution_steps} provides insight into the agent's reasoning complexity. As shown in \autoref{tb:prompts}, every ServiceNow AI Action accesses some incident and therefore the MCP tool for accessing the content of an incident is the most used tool, logging a total of over 2500 uses (see \autoref{fig:tool_usage}). The next most popular MCP tool is for querying Confluence documents for NOC procedures, which is used over 1000 times. Besides these commonly used tools, a variety of other MCP tools are used. This actual list of MCP tools is longer than the logical list shown in earlier discussions. One complication we observed with a long list of tools is that the AI model might have a hard time selecting the right one to use (see a discussion in \autoref{sec:skills}), which creates a situation where a request may need multiple steps alternating between language model processing and tool execution to find the right tools to gather information from multiple sources and synthesize the right information to complete the request.

\begin{figure}[htbp]
    \centering
    \includegraphics[width=0.6\columnwidth]{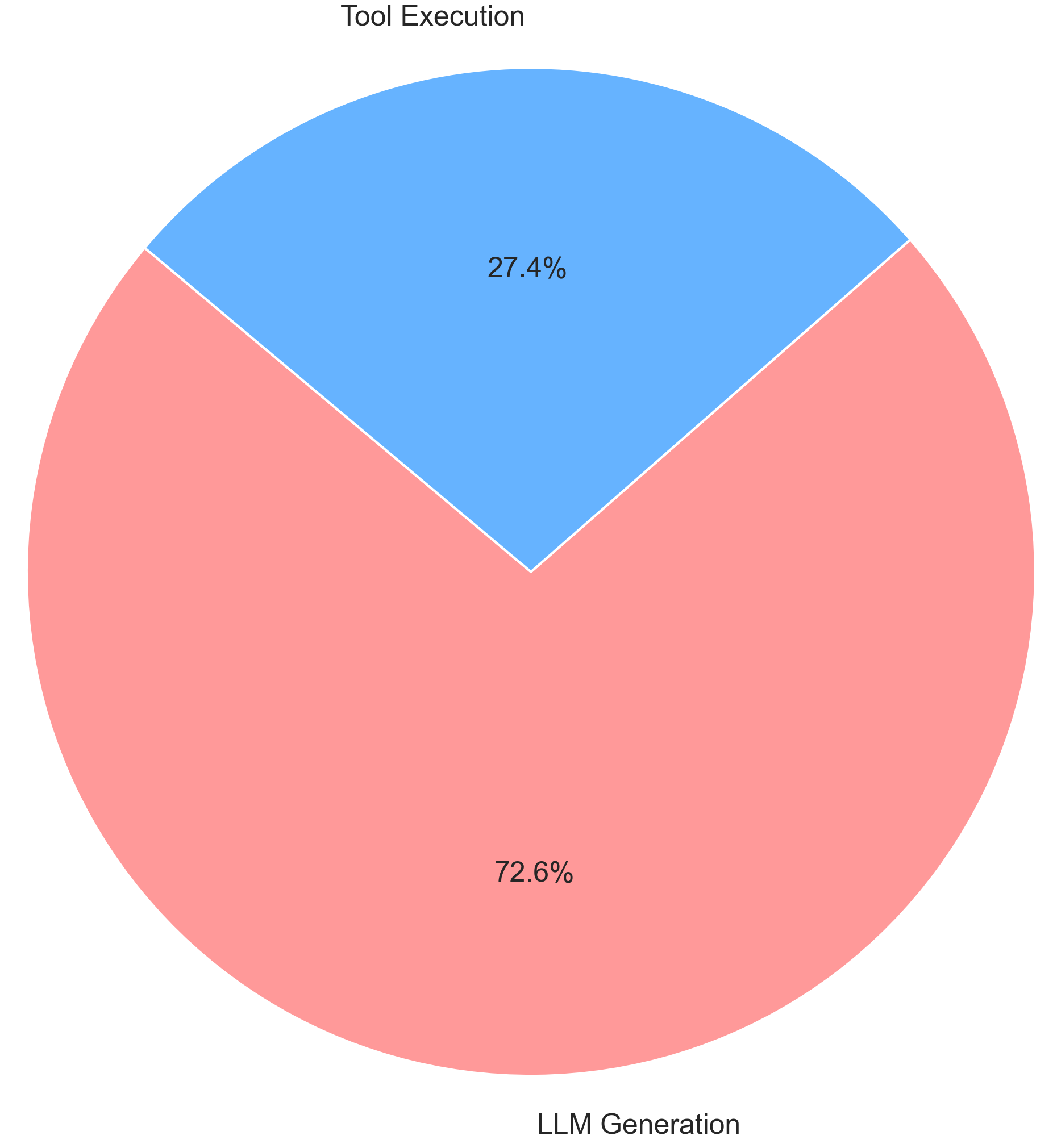}
    \caption{Overall time spent comparing large language model (LLM) generation versus external tool execution on the ORBIT Server.}
    \label{fig:time_breakdown}
\end{figure}

\autoref{fig:time_breakdown} contrasts the total time spent on language model generation versus external tool execution. Overall, 27.4\% of total execution time is spent on calling MCP tools for data access, while 72.6\% of the time is spent on invoking AI models. This suggests that we should investigate the time spent on AI models to understand how this might be optimized to reduce both the execution time and monetary cost. 

\subsection{Chat Interface Usage} \label{sec:chat-stats}

\begin{figure*}
  \centering
  \includegraphics[width=0.95\textwidth]{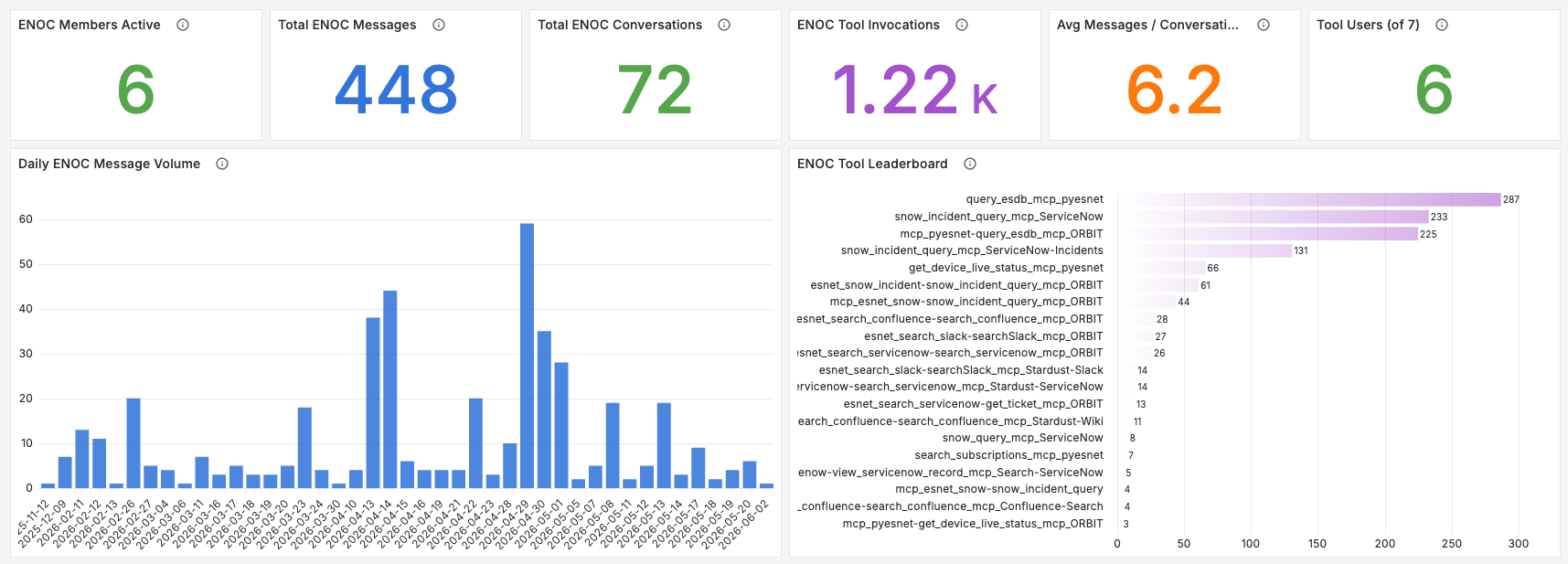}
  \caption{\texttt{chat.es.net} usage by NOC operators by June 2026. Six
    of seven members were active, exchanging 448 messages across 72
    conversations and invoking ORBIT-relevant tools (ESDB, ServiceNow,
    semantic search over Confluence and Slack, Stardust) 1,220 times without
    prescribed workflows.}
  \label{fig:enoc-chats}
\end{figure*}

The chat interface is used by the ORBIT team for tasks such as exploring how
to construct suitable prompt templates, and by NOC engineers for ad hoc
exploration of tickets and ticket-handling procedures
(i.e., Task \#6 in \autoref{tb:tasks}). \autoref{fig:enoc-chats} shows the
usage statistics of NOC operators through June. The LibreChat system
captured 72 conversations over the same period in which ServiceNow recorded
only 12 UI Actions, as shown in \autoref{tb:acts-by-user}. This disparity
can be partly explained by the fact that ServiceNow AI Actions cover only a
handful of tasks, while the chat interface gives operators much more freedom
to explore AI capabilities. Regardless, we observe considerably more
interest in the chat interface than in the ServiceNow UI Actions.

These patterns suggest that \texttt{chat.es.net} has value beyond its role
in ORBIT. Its flexibility makes it a natural general-purpose AI workbench
for ESnet staff, and the organic adoption observed among NOC operators
supports retaining it as a standalone service independent of the ORBIT
project's outcome.

\subsection{AI Model Usage} \label{sec:model-uses}

\begin{figure*}
    \centering
    \includegraphics[width=0.8\textwidth]{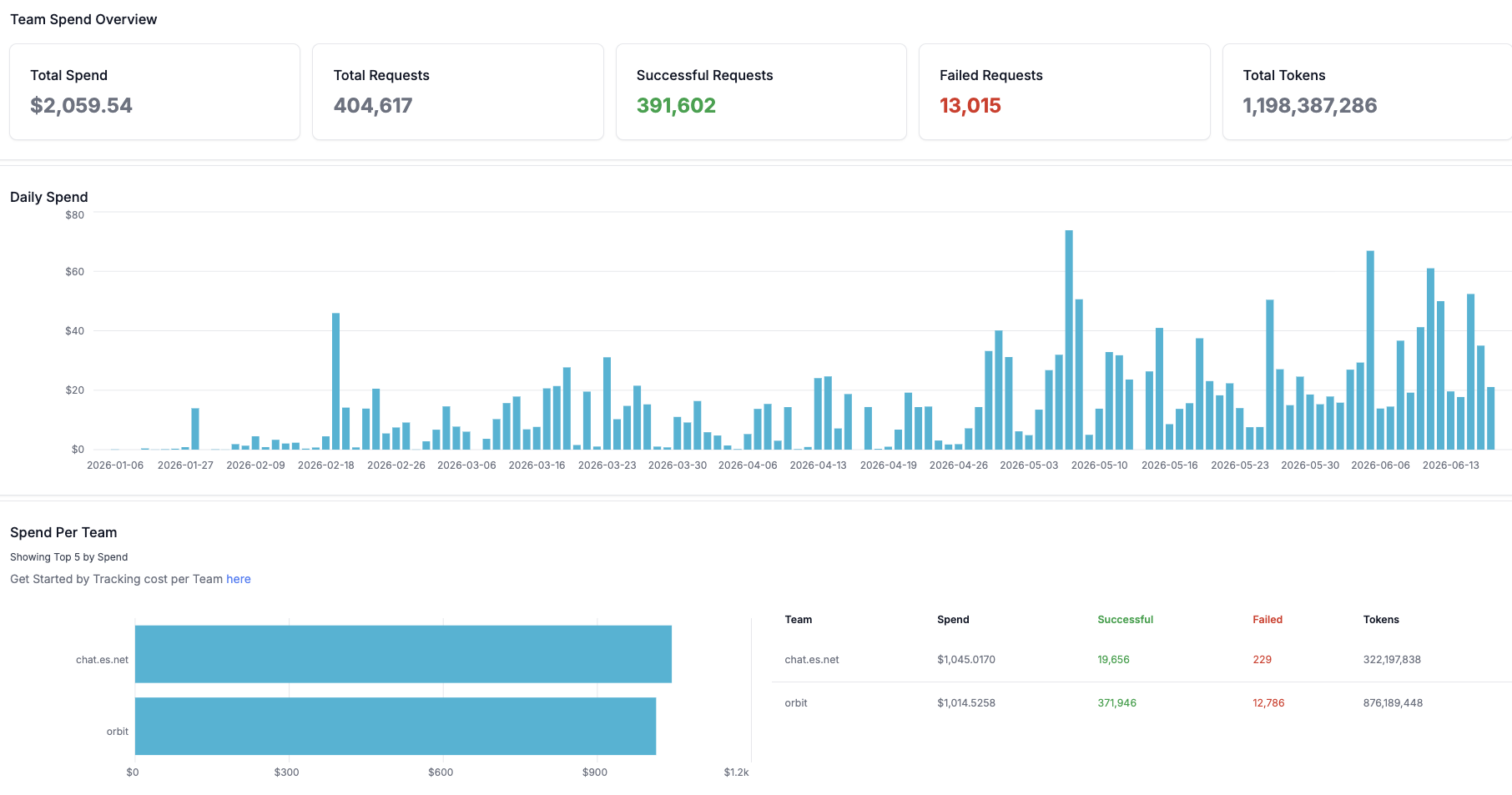}
    \caption{LiteLLM statistics about usage and cost from the
      \texttt{chat.es.net} and the project ORBIT.  Together these two
      services served just over 404,000 requests using about 1 billion
      tokens.  During the period of time, this LiteLLM server has served
      more than 714,000 requests consuming over 21 billion tokens.}
    \label{fig:cost}
\end{figure*}

The LiteLLM gateway described in this section realizes the AI Sandbox
(WP19) called for in ESnet's Data and AI report~\cite{Guok:2025:ESnetData}.
By providing a single authenticated entry point for model access, usage
accounting, and access control, it enables the organization to evaluate
demand for shared AI infrastructure across multiple workload types.

\autoref{fig:cost} is a screenshot of our LiteLLM installation showing the
cost and usage of requests going through the chat interface (i.e., the
component labeled~(3) in \autoref{fig:logical-view}) and the agentic
service behind ORBIT (i.e., the component labeled~(2) in
\autoref{fig:logical-view}).  The time period covered is the whole
development period.  \autoref{fig:cost} shows that the agentic service
consumed about 876 million tokens while the chat interface consumed about
322 million tokens.  Despite the difference in tokens consumed, their costs
in dollars are nearly the same.  This indicates that it is worthwhile to
select the right models to control costs for the enterprise.

\begin{table}[tb]
  \centering
  \caption{LiteLLM usage by workload category over the evaluation period.
    Coding assistants dominate token consumption despite fewer requests than
    chat and agentic workloads combined.}
  \label{tb:litellm-usage}
  \begin{tabular}{lrrr} \hline
    \textbf{Category} & \textbf{Requests} & \textbf{Tokens} & \textbf{Cost} \\
    & (000) & (000,000) &  (\$) \\ \hline
    Chat (chat.es.net)     &  19 & 322  & 1,045 \\
    Agentic (ORBIT)        & 372 & 876  & 1,014 \\
    Coding assistants      & 323 & 20,244 & 20,516 \\
    \textbf{Total}         & 714 & 21,422  & 22,620 \\ \hline
  \end{tabular}
\end{table}

\autoref{tb:litellm-usage} breaks down the LiteLLM usage by workload
category.  A little over 56\% of the requests going through the LiteLLM
server come from the chat service and the agentic service, yet these
requests consumed less than 5\% of the total tokens, which indicates that
the remaining workloads consume far more tokens per request.  Our inquiry
into this revealed that most of these requests originate from code
assistants that often send complex context for code refactoring and
generation.

This initial AI Sandbox deployment has attracted considerable attention from
ESnet staff, with several engineers reporting perceived productivity
improvements in both coding and operational tasks
(see also Section~\ref{sec:interviews}).  ESnet management has reviewed the usage statistics and authorized a \$500 per month cap for each ESnet staff.

\subsection{User Interviews} \label{sec:interviews}
During the evaluation phase of the ORBIT work, we conducted structured
interviews with our NOC engineers to assess system usability and gather
operational feedback.  The interviews produced two categories of finding:
observations about what operators experienced using the current system,
and directions for improvement that operators identified as high-value
next steps.  Both categories are summarized below.

The overarching usability finding is consistent across all interviews:
operators experience AI as most valuable when it reduces the effort of
stitching together cross-system incident context into a coherent,
actionable picture.  Adoption patterns varied by experience level, but
the core value proposition---reducing the cognitive overhead of
multi-source context assembly---was recognized uniformly.

\paragraph{Observations about the current system.}
\begin{itemize}

    \item \textbf{Context stitching is the dominant pain point.}
      Operators must manually reconstruct incident history from multiple
      disconnected tools (ServiceNow, Slack, email, and various
      monitoring systems).  ORBIT's cross-source aggregation directly
      addresses this, and operators recognized it as the feature with
      the greatest immediate impact.

    \item \textbf{Experience level shapes what ``assistance'' means.}
      Advanced engineers used AI sparingly but found value in targeted
      analytical tasks (cause-by-change analysis, incident updates,
      metric reviews), while less experienced users valued scaffolding
      such as command syntax hints and procedural reminders.  Both
      groups found the system useful, but for different task types.

    \item \textbf{Discoverability of data sources is a usability
      bottleneck.}  Users were often unaware that certain data
      (e.g., email context) was already accessible via existing MCP
      pathways, and they selected among MCP tools without a clear
      mental model.  This friction reduced the realized value of
      capabilities that were technically available.

    \item \textbf{Evidence-grounded, actionable outputs are valued.}
      Users cited concrete value when AI produced a correct,
      operationally actionable insight (e.g., identifying that a peer
      had recently turned up a relevant circuit).  Generic summaries
      without operational grounding were less appreciated.

\end{itemize}

\paragraph{Directions for improvement identified by operators.}
\begin{itemize}

    \item \textbf{Workload-level roll-ups, not only per-ticket
      assistance.}  A recurring unmet need was the ability to summarize
      all active incidents (e.g., 15--30 items) with provider updates
      and next actions, particularly for shift handoff.  This points to
      a workload-level summarization capability beyond the current
      per-ticket scope.

    \item \textbf{Better handling of repeated and recurring incidents.}
      When tickets reappear without root-cause closure, operators
      re-investigate from scratch because historical resolution context
      is not aggregated into an actionable form.  Operators requested
      that ORBIT surface prior resolution attempts for recurring
      incident patterns.

    \item \textbf{Consolidation of alarming systems.}  Operators
      repeatedly requested a unified conversational interface across
      DNA, Spectrum, Alerta, and TNMS.  Integrating additional alarming
      data sources would be the highest-leverage next integration step.

    \item \textbf{Catch-up and notification triage.}  Operators
      described recurring ``what did I miss'' workflows across
      Slack and email, and flagged noise from non-operational
      notifications.  A digest and triage capability would reduce
      daily overhead considerably.

    \item \textbf{Externalization of institutional knowledge.}  Key
      knowledge---CLI syntax quirks, topology and vendor specifics
      such as satellite-based circuit details---was not reliably
      accessible from static documentation.  Operators requested that
      ORBIT's retrieval layer be extended to surface this tacit
      knowledge more reliably.

\end{itemize}

\section{Lessons and Discussion} \label{sec:lessons}

\subsection{From Prompts to Skills: Managing Complexity in AI Tool Chains} \label{sec:skills}

The current industry best-practice to deal with the stochastic behavior in AI tool chain is to emply a technique known as ``skills'' which is to provide a common set of directives about how to use a specific set of MCP tools to in an application context such as ORBIT.  A central engineering insight from ORBIT development is that prompts must be treated with the same rigor as source code: versioned, tested, and iteratively refined.  The difference between a marginally useful AI response and a genuinely operational one is almost always traceable to prompt engineering rather than model selection.  

To illustrate, Table~\ref{tab:prompt-comparison} contrasts a na\"ive prompt with an engineered, stepwise variant for the same incident-summarization task.  The na\"ive prompt (``Please summarize this ticket with comments and work notes'') required 10~agent actions, 78\,s to complete, and 3~retries before the agent located the correct query path.  The engineered prompt---which explicitly instructs the agent to retrieve the \texttt{sys\_id} first, specifies the fields of interest, and directs parallel queries---completed in 4~agent actions and 57\,s (a 26.6\% reduction in wall-clock time) with correct execution on the first attempt.  The existing experiment described in Section 6.1 is to refine the instruction for a specific task.  Based on the industrial best-practice, we are in the process of developing ``skills'' for ORBIT so that shared knowledge such as which tool to use to turn a ticket number into text version of the ticket content is always available does not need to be write into the prompt template for any individual task.  

\begin{table}[htbp]
\centering
\caption{Comparison of na\"ive versus engineered prompts for incident
  summarization.   Note that the engineered prompt is about 26.6\% faster.}
\label{tab:prompt-comparison}
\begin{tabular}{lcc}
\hline
& \textbf{Na\"ive} & \textbf{Engineered} \\
\textbf{Metric} & \textbf{Prompt} & \textbf{Prompt} \\
\hline
Agent actions & 10 & 4 \\
Completion time & 78\,s & 57\,s \\
Retries to correct query & 3 & 0 \\
\hline
\end{tabular}
\end{table}

This result reinforces a practical guideline: version your prompts, test them against representative incidents, and iterate systematically.  Within ORBIT, prompts are maintained as versioned artifacts alongside the agent code, and prompt variants are evaluated using controlled incident replays (Section~5) so that regressions are caught before deployment.

\subsection{Managing Stochasticity: Same Prompt, Different Answer}

AI models are inherently stochastic---randomness in token sampling is a feature of the architecture, not a bug.  In practice, this means that the same prompt applied to the same incident can produce subtly (or not-so-subtly) different outputs on successive runs.  Two additional sources of non-determinism compound the problem in agentic systems such as ORBIT:

\begin{itemize}
    \item \textbf{Tool-selection ambiguity.} When MCP tool names or descriptions are similar, the model occasionally selects the wrong tool, leading to divergent execution paths even when the prompt is identical.
    \item \textbf{Retrieval variability.} Semantic search via RAG introduces its own layer of non-determinism: slight differences in embedding scores or index state can alter which documents are surfaced and, consequently, the content of the generated response.
\end{itemize}

Several practices help mitigate---though never fully eliminate---this variability:

\begin{itemize}
    \item \textbf{Consistent, descriptive tool naming with documentation.} Clear, unambiguous MCP tool names and rich tool descriptions reduce the likelihood of mis-selection by the reasoning model.
    \item \textbf{Richer prompts that guide tool selection.} Stepwise instructions that explicitly name the tools to invoke (as in Lesson~1) narrow the model's decision space and improve first-try correctness.
    \item \textbf{Ongoing refinement of templates and model selection per task type.} Different tasks may benefit from different models or temperature settings; matching these systematically reduces output variance for high-stakes actions.
\end{itemize}

The ORBIT team is adopting the industry best practice of encapsulating these mitigation strategies into reusable ``skills.''  These skills provide a structured way to manage the stochastic nature of AI tool chains by providing a consistent set of instructions for the AI to follow.  
Stochasticity never fully disappears.  The practical takeaway is to design around it: use deterministic anchors (explicit tool routing, structured output schemas, and retry-with-validation logic) wherever possible, and instrument the system to detect when outputs diverge beyond acceptable bounds.

\subsection{Emergent Use Cases: The Best Ones Find You}
\label{sec:usecases}

Not all high-impact use cases can be anticipated during project planning.  One of ORBIT's most effective deployments emerged organically from an existing operational ritual: the weekly metrics meeting.

Each week, the NOC reviews incidents that fall outside service-level specifications.  A recurring analysis task during this meeting is identifying all incidents that resulted from a planned change and then verifying two conditions in real time: (1)~whether the associated maintenance window was exceeded, and (2)~whether all affected configuration items (CIs) were included in the change record.  Performing this cross-referencing manually during a live meeting was slow and error-prone.

A NOC engineer collaborated with the Business Automation team to add a ServiceNow UI action that automates this analysis using ORBIT's AI infrastructure.  The implementation took approximately one hour---enabled by the fact that the underlying platform, data integrations, and prompt patterns were already in place.

The broader lesson is architectural: \emph{build the platform, then watch what your users do with it.} By investing in composable data integrations and a flexible prompt/tool framework, the team created the conditions for rapid, user-driven innovation.  The highest-impact use case may not be the one originally planned; it may instead surface from operators who recognize a painful workflow that the platform can now address with minimal additional effort.  This experience also led to the ``Caused-By-Change Analysis'' and ``Change Analysis'' actions visible in Table~\ref{tb:acts-by-task}, which were not part of the original six-task scope.

\subsection{Organic Infrastructure Adoption: Build It and They Will Come} \label{sec:changeAnalysis}
A recurring theme in AI deployment guidance is that infrastructure
investments should be justified by projected demand.  ORBIT's experience
suggests the opposite sequence: build shared infrastructure for a concrete
use case, instrument it, and let demand reveal itself.

The LiteLLM gateway (\autoref{sec:litellm}) was deployed to serve ORBIT's
agentic service and the chat interface -- two workloads with a combined
1.2~billion tokens over the evaluation period (\autoref{tb:litellm-usage}).
Without additional promotion, the same gateway attracted coding-assistant
workloads that consumed over 20 billion tokens -- roughly 95\% of total
token volume from engineers who discovered the endpoint through word of
mouth.  Similarly, the chat interface (\autoref{sec:chat-stats}) saw 6 of 7
NOC operators adopt it organically, generating 72 conversations and over
1,200 tool invocations (\autoref{fig:enoc-chats}), compared with only 12
user-initiated ServiceNow AI Actions in the same period
(\autoref{tb:acts-by-user}).  In both cases, the heaviest usage came from
workloads and interaction patterns that were not part of the original
project scope.

Two practical implications follow.  First, shared AI infrastructure,
including model gateways, authenticated tool endpoints, usage accounting,
should be designed from the outset as general-purpose services rather than
project-specific utilities; the marginal cost of supporting additional
workloads is low once the platform exists, and usage data
(\autoref{fig:cost}) provides the evidence base for sustaining the
investment.  Second, per-workload cost visibility matters: the data in
\autoref{tb:litellm-usage} show that coding assistants and agentic services
have dramatically different token-per-request profiles, and therefore
different cost structures; without workload-level accounting, the
organization cannot make informed model-selection or budget decisions as
adoption scales.

\subsection{Cognitive Barriers to AI Adoption}

Perhaps the most unexpected lesson was not technical but organizational.  AI is sufficiently new in operational settings that most staff have difficulty envisioning concrete use cases.  The prevailing mental model treats AI as a glorified chatbot, and few practitioners are tracking the rapid evolution of agentic capabilities, tool integration, and retrieval-augmented generation.

This creates a cognitive barrier: moving from \emph{constraint-based thinking} (``what can this chatbot do?'') to \emph{possibility-based thinking} (``what workflows could an AI agent with access to our data sources transform?'') is genuinely difficult without concrete, relatable examples.  However, once operators were given a hands-on demonstration of what ORBIT could do with their own tickets and data---particularly the experiences described in Lessons~1--3---the ideation barrier dropped rapidly.  New use-case proposals began to flow from the NOC team itself, several of which were implemented within hours.

This dynamic carries several practical implications for teams deploying operational AI:

\begin{itemize}
    \item \textbf{Do not expect your initial use cases to be your strongest.} Early use cases serve as existence proofs and learning vehicles; the most impactful applications typically emerge after the team has internalized what the platform can do.
    \item \textbf{Infrastructure begets opportunity.} The more data sources are integrated and the more tools are exposed, the larger the combinatorial space of feasible use cases becomes.  Each new MCP server or indexed knowledge base multiplies the options available to both the agent and the human designers.
    \item \textbf{Embrace rapid iteration.} AI development rewards fast cycles of prompt engineering, testing, and deployment.  Teams should be organizationally prepared for this tempo, including lightweight approval processes for new AI actions and prompt variants.
    \item \textbf{Develop organizational policies early.} Clear policies for AI usage---covering data handling, model selection, output review, and accountability---provide the guardrails that enable experimentation without unacceptable risk.
    \item \textbf{Focus on security, but do not overcorrect.} Overly restrictive policies can stifle the exploratory culture that makes operational AI successful.  The goal is to establish security boundaries (e.g., write-scope restrictions, audit logging, role-based access) that protect sensitive data while preserving the freedom to iterate on prompts, tools, and workflows.
\end{itemize}

\section{Review of the State-of-the-art} \label{sec:related}

AI/ML is widely expected to assist Network
Operations~\cite{Bobie2024network, Min2024AIops}.  However, widely
publicized studies document the difficulty of transferring successes from
demonstration projects to operational
environments~\cite{Rajan2025pilots, Shaik2026AI, Vallone2025AI}.  It is
therefore highly desirable to ground demonstrations in realistic business
use cases, incorporating proper business processes and engaging the right
staff~\cite{Shaik2026AI}.  This report primarily focuses on the technical
aspects of the ORBIT work, but also touches on data governance and staff
engagement.

There are many publications showing successful uses of AI technology in various use cases~\cite{Joy2024AIops}.  Here we briefly summarize one specifically focuses on network operations in cloud infrastructure~\cite{Veluru:2021:Incident}.  A featured case study demonstrates the tangible impact on network operations, showing a 60\% reduction in false positives and a decrease in Mean Time to Recovery (MTTR) from three hours to 55 minutes.

In many of these studies, a "human-in-the-loop" approach is needed to ensure operational safety and trust~\cite{besigomwe2025human, Ottun:2025:Trustworthy, Tariq:2025:AlertFatigue}.  In a network operations case study, site reliability engineers validated AI-suggested remedies and restricted auto-remediation to high-confidence predictions~\cite{Veluru:2021:Incident} to ensure operational safety, prevent "black box" errors, and leverage expert feedback to refine models.

\paragraph{Retrieval-Augmented Generation (RAG) for Enterprise Knowledge}
Retrieval-Augmented Generation (RAG) is a technique that enhances AI language model responses by dynamically fetching relevant, up-to-date information from external knowledge sources at query time---grounding the model's outputs in real data rather than relying solely on what was learned during training~\cite{Lewis:2020:RAG}.  While this process reduces hallucinations and improves accuracy, particularly in specialized technical domains, it also introduces significant challenges~\cite{Wang:2024:RAG}.  Key among these are ensuring the model remains robustly grounded, implementing reliable citation and attribution, and addressing limitations when retrieved evidence is incomplete, conflicting, or must be synthesized across multiple sources~\cite{Yu:2025:Multi-RAG}.

\paragraph{Semantic Search over Siloed Operational Data}
Network operations at ESnet generate vast amounts of heterogeneous data across disparate systems like ticketing platforms, wikis, and chat logs.  This proliferation of data silos hampers efficient knowledge discovery and decision-making, leaving critical information underutilized and unsuitable for integrated querying~\cite{nagabhyru2023silos}.  To address this, there is a pressing need to not only break down these enterprise data silos but also to provide a simple, unified query interface~\cite{Masmoudi:2024:Semantic}.  Such an interface requires robust semantic query understanding to accommodate the natural variations in ESnet's operational workflows—seamlessly interpreting synonyms, technical jargon, and identifier-based searches to provide operators with a single, coherent view of all relevant institutional knowledge.

\paragraph{Hybrid Retrieval Architectures and Rank Fusion}
To overcome the limitations of any single retrieval method, our approach utilizes a hybrid architecture that combines the complementary strengths of lexical (e.g., BM25), dense, and sparse retrieval models~\cite{Cormack:2009:RRF}.  The resulting ranked lists are merged using rank fusion, with methods like Reciprocal Rank Fusion (RRF) being particularly effective\cite{Cormack:2009:RRF, Bruch:2023:RankFusion}.  RRF enhances overall robustness by rewarding documents that rank well across multiple retrievers — even if no single retriever ranks them first — mitigating the weaknesses of any individual approach.  This hybrid system is further augmented by a high-precision strategy for exact identifier matching, which handles the critical edge case of structured, known-item queries common in operational data.

\paragraph{Prompting and Orchestrating LLM Tool Use (``Agentic'' Systems)}
The foundational research in LLM-based autonomous agents emerged from several seminal works: ReAct \cite{yao2023react} introduced the paradigm of interleaving reasoning traces with actions, enabling models to dynamically plan and adjust based on environmental feedback. AutoGPT \cite{autogpt2023} demonstrated fully autonomous task execution by chaining LLM calls with memory and tool use, sparking widespread interest in self-directed agents. Generative Agents \cite{park2023generative} showed that LLMs could simulate believable human behavior in sandbox environments through memory retrieval, reflection, and planning mechanisms. These early efforts were later unified by comprehensive surveys---Wang et al. \cite{wang2023survey} established a four-module framework (Profiling, Memory, Planning, Action), while Xi et al. \cite{xi2025rise} offered a cognitive Brain-Perception-Action paradigm that extended the discourse to multi-agent societies and human-agent collaboration.

\paragraph{Evaluation Methodologies for Operational AI} 
   Recent research and development in large language model (LLM)-based agent evaluation has shifted toward more comprehensive and dynamic assessment frameworks. Yehudai et al.~\cite{yehudai2025survey} provide a systematic review of agent benchmarking, categorizing evaluation approaches from atomic core capabilities such as planning and tool-use to application-specific and generalist assessments, while emphasizing the growing importance of continuously updated dynamic benchmarks to address the rapid evolution of agent capabilities. Complementing this, Luo et al.~\cite{luo2025agent} present a methodology-centered taxonomy synthesizing over 300 papers, linking architectural foundations to deployment challenges and practical tooling considerations. Domain-specific evaluation has also emerged as a critical area, exemplified by Sun et al.~\cite{sun2025data}, who survey ``Data Agents'' capable of autonomous statistical reasoning, code generation, and tool-augmented retrieval for data science pipelines. Together, these works highlight a maturing field that increasingly demands multi-dimensional evaluation spanning reasoning, tool integration, real-world applicability, and domain expertise.

\paragraph{Measuring Usability, Reliability, and Cost}
Serving AI models in an enterprise environment requires a multi-layered, Zero-Trust security posture that addresses threats from the infrastructure to the model interaction layer~\cite{Huijts2025AIGateway, Mao2025APIGateway, Reddy:2026:0TrustLLM}.  These approaches typically start by assuming the traditional network perimeter is obsolete, instead enforcing continuous verification of identities, devices, and workloads with adaptive, AI-driven access policies~\cite{Nangi2023:0TrustGateway, Reddy:2026:0TrustLLM}.  Critically, this extends to the LLM itself through a dedicated "Secure AI Gateway," which treats all incoming user prompts as untrusted~\cite{Brett2025secureMCP, Sharma2024securingLLM}.  Our current implementation uses LiteLLM for enforcing Role-Based Access Control (RBAC), preventing data exfiltration, and detecting adversarial attacks like prompt injection before they reach the model~\cite{LiteLLM}.

Beyond robust security, successful operational deployment hinges on comprehensive measurement and governance. This requires diligent instrumentation and telemetry to continuously evaluate system performance against key enterprise metrics: usability (e.g., task success, user satisfaction), reliability (e.g., timeouts, error rates), and cost (e.g., token consumption)~\cite{islam2023measuring, Mao2025APIGateway}.  Capturing this data enables reproducible evaluations, provides a clear accounting of the system's value and resource utilization, and establishes the foundation for effective data governance and iterative improvement in a production setting.

\paragraph{Distinctiveness of ORBIT}
The works surveyed above advance individual aspects of operational AI,
anomaly detection and incident triage, retrieval-augmented generation,
hybrid search, agentic orchestration, and evaluation methodology, but they
largely address these concerns in isolation and evaluate them on curated
benchmarks rather than in production operator workflows.  ORBIT targets the
gap between these demonstrated capabilities and their integrated, measurable
use in a real NOC setting.  Specifically, (i) it synthesizes evidence across
multiple siloed operational data sources (ServiceNow tickets, Confluence
procedures, Slack discussions, and network databases) within a single
agentic action, rather than retrieving from a single corpus; (ii) it embeds
AI outputs directly into the existing ServiceNow incident workflow so that
operator acceptance, editing, and rejection provide naturalistic evaluation
signals without additional tooling; (iii) it treats prompts as versioned,
testable engineering artifacts and evaluates prompt variants through
controlled incident replays, making prompt optimization an explicit part of
the system evaluation rather than an offline exercise; and (iv) it reports
both system-level performance metrics (request latency, token consumption,
tool-call depth) and operator-level usability evidence (structured
interviews, adoption patterns, feedback signals) from a sustained
deployment, providing the kind of workflow-aligned evaluation that prior
work calls for but rarely demonstrates end to end.

\section{Conclusion and Future Work} \label{sec:conclusion}
This report presented ORBIT, an agentic AI system integrated into ESnet's
NOC incident workflow, and reported on the system design, operational
measurements, and engineering lessons gathered during its initial
deployment.

\subsection{Project Outcomes and Lessons Learned}
The project was organized around three objectives: build
software to accomplish six specific NOC tasks, engage ESnet staff in
exploring AI tools, and gather performance and usability statistics to
inform future AI efforts.  We revisit each in turn.

\paragraph{Software and task coverage}
ORBIT delivered all six originally scoped NOC tasks
(\autoref{tb:tasks}), and the platform's composability enabled two
additional tasks---Caused-By-Change Analysis and Change
Analysis---proposed by NOC engineers and implemented within hours
(see Section~\ref{sec:changeAnalysis} for details, and \autoref{tb:acts-by-task}
for invocation counts).  The system recorded 169~AI Actions in
ServiceNow over the evaluation period, of which nearly 70\% were
triggered automatically by business rules (\autoref{tb:acts-by-user}).
Recommendation tasks (suggest alarm procedure, recommend alarm priority)
were invoked most frequently but exhibited high variance in execution
time, with standard deviations exceeding their means; summarization tasks
showed more predictable performance (\autoref{tb:acts-by-task}).  These
patterns confirm that cross-source synthesis tasks are both the most
demanded and the most sensitive to prompt and tool-selection quality---a
finding reinforced by the skill-refinement results in
Section~\ref{sec:lessons}, where engineered skills reduced agent actions
from 10 to~4 and eliminated retries (Table~\ref{tab:prompt-comparison}).

Output quality is supported by a systematic continuous monitoring regime.
Drawing on NOC engineer review of more than 200~ServiceNow AI action
instances, the team distilled a regression suite of approximately four
dozen tests executed daily against production-representative tickets.
This provides longitudinal evidence that the system maintains consistent
behavior across prompt updates and model changes, and that regressions
are detected before they reach operators
(Section~\ref{sec:server-stats}).

\paragraph{Staff engagement}
Adoption data show organic uptake beyond the original project scope.
Six of seven NOC operators used the chat interface, generating
72~conversations and over 1,200 tool invocations without prescribed
workflows (\autoref{fig:enoc-chats})---substantially more interaction
than the 12~user-initiated ServiceNow AI Actions recorded in the same
period (\autoref{tb:acts-by-user}).  The LiteLLM gateway attracted
coding-assistant workloads that consumed over 20~billion tokens---roughly
95\% of total token volume---from engineers who were not part of the
ORBIT project (\autoref{tb:litellm-usage}).  User interviews
(Section~\ref{sec:interviews}) identified cross-source context
aggregation as the primary value of AI assistance and highlighted
tool discoverability as the main usability bottleneck: operators were
often unaware which data sources were already reachable and selected
among MCP tools without a clear mental model.

\paragraph{Performance and cost visibility}
The instrumentation built into ORBIT---execution traces, per-step
timing, token accounting, and the LiteLLM usage dashboard---provided
the quantitative foundation for all evaluation results reported in
Section~\ref{sec:perf}.  A practical finding is that workload-level cost
visibility is essential: the agentic service and chat interface together
consumed less than 5\% of a project's total tokens, yet accounted for over 56\% of
requests, while coding assistants dominated token consumption at
dramatically different cost-per-request profiles
(\autoref{tb:litellm-usage}).  Without per-workload accounting, the
organization could not make informed model-selection or budget decisions
as adoption scales.

\paragraph{Lessons and implications}
Five engineering lessons emerged from the deployment
(see~\autoref{sec:lessons}).  The ORBIT team has begun to adopt the industry
best practice of encapsulating procedural knowledge into versioned, testable
``skills''; this approach helps mitigate the inherent stochasticity of AI
toolchains.  The team also found that the highest-impact use cases may
emerge from operators after the platform is available rather than from
initial planning, and that staff adoption accelerates sharply once operators
see concrete demonstrations with their own data.  Together, these lessons
argue for investing in composable, general-purpose AI infrastructure---model
gateways, authenticated tool endpoints, usage accounting---and then
instrumenting it to let demand and use-case evolution guide subsequent
development.

\subsection{Recommendations}
Based on the evidence gathered, we offer the following recommendations for
next steps with AI infrastructure and NOC tooling.

The four general-purpose components introduced by ORBIT: the LiteLLM
model gateway, the agentic application service, the \texttt{chat.es.net}
interface, and the MCP tool layer, have demonstrated sustained demand well
beyond the original project scope.  The LiteLLM gateway served over
714,000~requests and 21~billion tokens, with 95\% of that volume coming
from engineering workloads outside ORBIT; six of seven NOC operators
adopted the chat interface organically; and the MCP tool layer enabled
rapid implementation of new use cases proposed by operators.  We therefore
recommend that ESnet formally adopt these four components as officially
supported organizational services, with dedicated operational ownership,
documented SLAs, and integration into ESnet's standard access-control and
cost-accounting infrastructure.

The ServiceNow AI Actions, by contrast, recorded only 12 user-initiated
invocations from NOC operators during the evaluation period, with most
automated actions attributable to business-rule triggers rather than active
operator engagement.  Output quality evaluation is still in progress, and
the evaluation period predates several prompt optimizations.  We therefore
recommend that a decision on the long-term status of the AI Actions be
deferred pending the structured quality evaluation and extended usage
collection planned for the coming months~(Section~\ref{sec:aiactions}).

Finally, transitioning any of these components from exploratory project
infrastructure to officially supported services requires careful
productization planning.  This includes establishing clear operational
ownership and on-call responsibilities, defining incident-response
procedures for service degradation, documenting dependency chains
(particularly the LiteLLM gateway's dependencies on external model
providers), and setting up capacity planning aligned with observed
growth in token consumption.  The engineering lessons in
Section~\ref{sec:lessons} provide a starting point, but a dedicated
productization plan---covering staffing, governance, security review,
and user-facing SLAs---should be developed before any component is
formally promoted to production service status.

\subsection{Future work}
The following items address both the technical evolution of the ORBIT
platform and the strategic business considerations that will determine
whether its infrastructure can be sustained and expanded responsibly.

\emph{Systematic Evaluation of Mission-Oriented Outcomes.} The current
evaluation rightly focuses on system performance and usage metrics, but a
critical missing piece is a formal, systematic evaluation of the quality and
accuracy of the AI-generated outputs, and its impact on ESnet's operational
mission.  Without this, the true operational value and potential risks of
ORBIT remain unquantified.  Future evaluation cycles must focus on
understanding the low adoption of the ServiceNow AI Actions and measure
outcomes that connect directly to ESnet's network reliability mission:
operator time-to-context before and after ORBIT assistance, reduction in
repeated investigative effort for recurring incidents, and shift-handoff
quality as assessed by the receiving operator.  This framework is a
prerequisite for moving ORBIT from a promising prototype to a trusted,
production-grade operational tool.

\emph{Development of a Skill Library.}  To further mitigate the impact of
stochasticity and to ensure the consistent application of best practices,
future work should focus on the development of a comprehensive skill
library.  This library will contain a collection of versioned, tested, and
documented skills for each of the NOC tasks that ORBIT supports.  This will
not only improve the reliability and predictability of the system, but will
also serve as a valuable resource for training new staff and for sharing
knowledge across the organization.

\emph{Unified identity-aware tool gateway.}  A recurring theme across the
system design and operator feedback is that ORBIT's value depends less on
any single model or tool than on how cleanly its tools are exposed,
discovered, and governed.  ORBIT currently spans three distinct tool-access
paths---in-process, gateway-brokered, and identity-gated
(Section~\ref{sec:mcptools})---that trade off latency, centralized
governance, and end-user identity propagation differently, with no single
path resolving all three concerns simultaneously.  A natural next step is to
converge these paths onto a unifying, identity-aware tool gateway: a single
brokerage point through which tools are discovered, governed, and invoked,
and through which the authenticated user's identity propagates to the
data-owning service so that authorization decisions are made against the
real end user rather than a shared service credential.  The relevant
building blocks are emerging---OIDC token exchange (RFC~8693) for delegating
identity across service hops, and the evolving MCP authorization
specification---but the standards remain young and reference implementations
are still maturing.

\emph{Operational sustainability and knowledge concentration.}  The depth of
knowledge required to maintain, prompt-optimize, and extend ORBIT currently
resides in a small development team.  If key personnel leave or shift to
other projects, operational continuity is at risk.  Future work should
include structured knowledge transfer: documented runbooks for common
failure modes, a maintained library of annotated prompt examples covering
each task type, and cross-training of at least one NOC-side maintainer on
prompt development and regression testing workflows.  This is a prerequisite
for the formal service adoption recommended above.

\emph{Data Governance and Provider Dependency.} ORBIT processes live
ServiceNow incident records and ESDB network configuration data.  While the
LiteLLM gateway routes commercial requests through FedRAMP-compliant
infrastructure, there is currently no formal classification policy governing
which data categories may be sent to external model providers and under what
contractual terms.  Future work should establish a data classification
scheme aligned with ESnet's information security policy, implement prompt-
and response-level filtering to prevent sensitive fields (e.g., device
credentials, circuit topology) from being included in external model calls,
and audit existing AI Action prompts against that classification.
Furthermore, many AI workloads currently route through commercial model
providers, which could be an operational risk.  Future work should establish
a systematic model-equivalence evaluation framework that enables workloads
to be migrated across providers---or to on-premises alternatives---with
measured quality impact, and should define the criteria under which
on-premises models become the default for cost-sensitive or
security-sensitive tasks.

\section*{Acknowledgment}
This work was supported in part by the Office of Advanced Scientific Computing Research, Office of Science, of the U.S. Department of Energy under Contract No.~DE-AC02-05CH11231, and used resources of the Energy Science Network (ESnet).

\bibliographystyle{abbrv}
\bibliography{ref,software}

\end{document}